\documentclass[preprint,prd,aps,showpacs,showkeys,nofootinbib]{revtex4}
\usepackage{graphicx}
\usepackage{dcolumn}
\usepackage{bm}
\begin{document}

\preprint{arXiv: 0805.0653}

\title{The two loop supersymmetric corrections to lepton
anomalous dipole moments in split supersymmetry scenarios}

\author{Tai-Fu Feng, 
Lin Sun, Xiu-Yi Yang}

\affiliation{Department of Physics, Dalian University of Technology,
Dalian, 116024, China}

\date{\today}

\begin{abstract}
An analysis of electroweak corrections to the anomalous dipole moments of lepton
from some special two-loop diagrams where a closed neutralino/chargino loop is
inserted into relevant one-loop diagrams of the standard model is presented
in the split supersymmetry scenarios. Considering the translational invariance of the inner
loop momenta and the electromagnetic gauge invariance, we get all dimension 6
operators and their coefficients. After applying equations of motion
to the external leptons, we obtain the anomalous dipole moments of lepton.
The numerical results imply that there is parameter space where the contribution
to the muon anomalous magnetic dipole moment from this sector is perhaps significant,
and the contribution to the electron electric dipole moment from this sector
is sizable enough to be observed in next generation experiments.
\end{abstract}


\maketitle

\section{Introduction \label{sec1}}
\indent\indent
At both aspects of experiment and theory, the magnetic dipole moment (MDM) of lepton
draws the great attention of physicists because of its obvious
importance. The anomalous dipole moments of lepton not only can be
used for testing loop effect in the standard model (SM), but also
provide a potential window to detect new physics beyond the SM.
The current experimental result of the muon MDM is \cite{exp}
\begin{eqnarray}
&&a_{_\mu}^{exp}=11\;659\;208\;\pm\;6\;\times 10^{-10}\;.
\label{data}
\end{eqnarray}

From the theoretical point of view, contributions to the muon MDM are generally
divided into three sectors \cite{Jegerlehner}: QED loops, hadronic contributions
and electroweak  corrections. With the hadronic contributions
which are derived from the most recent $e^+e^-$ data, we can get
the following SM predictions \cite{sm1,sm2,sm3}:
\begin{eqnarray}
&&a_{_\mu}^{SM}=11\;659\;180.9\;\pm\;8.0\;\times 10^{-10}\;,
\nonumber\\
&&a_{_\mu}^{SM}=11\;659\;175.6\;\pm\;7.5\;\times 10^{-10}\;,
\nonumber\\
&&a_{_\mu}^{SM}=11\;659\;179.4\;\pm\;9.3\;\times 10^{-10}\;.
\label{sm}
\end{eqnarray}
The deviations between the above theoretical predictions and the
experimental data are all approximately within error range of
$\sim2\sigma$. Although this $\sim2\sigma$ deviation cannot be
regarded as strong evidence for new physics, along with the
experimental measurement precision and theoretical prediction
accuracy being constantly improved, this deviation may become more
significant in near future.

In fact, the current experimental precision ($6\times 10^{-10}$)
already puts very restrictive bounds on new physics scenarios.
In the SM, the electroweak one- and two-loop contributions amount to $19.5\times
10^{-10}$ and $-4.4\times10^{-10}$ \cite{sm-2l} respectively.
Comparing with the standard electroweak corrections, the electroweak
corrections  from new physics are generally suppressed by $\Lambda_{_{\rm EW}}^2/\Lambda_{_{\rm NP}}^2$,
where $\Lambda_{_{\rm EW}}$ denotes the electroweak energy scale and
$\Lambda_{_{\rm NP}}$ denotes the energy scale of new physics.

Supersymmetry (SUSY) has been considered a most prospective candidate for
new physics beyond the SM. Nevertheless, the softly broken
SUSY at electroweak scale induces many unwanted phenomenological problems,
such as new sources of flavor changing neutral currents, new CP violating
phases etc. In order to solve those problems, the authors of literature \cite{Arkani1}
have recently proposed a split scenario.
In this split scenario, SUSY is broken at a high energy scale that could be
even near the scale of grand unification theory (GUT). A direct result
of this assumption is that the scalar superpartners of SM fermions are
all super heavy. On the other hand, charginos and neutralinos acquire masses
around electroweak scale to TeV or so because of R-symmetry and PQ symmetry.
Within this framework, heavy sfermions suppress the one-loop supersymmetric
corrections to the processes of flavor changing neutral currents and lepton
anomalous dipole moments at a negligible level. The leading contributions of new
physics to theoretical predictions only arise from the two-loop diagrams
in which a closed neutralino/chargino loop is inserted into relevant
one loop SM diagrams, and the corresponding theoretical corrections are sizable enough to be well
within the sensitivity of the next generation of experiments \cite{Arkani2}.

Actually, the two-loop electroweak corrections to the anomalous
dipole moments of lepton are discussed extensively in literature.
Utilizing the heavy mass expansion approximation (HME) together
with the corresponding projection operator method, Ref.\cite{czarnecki}
has evaluated the two-loop standard electroweak corrections to the muon MDM.
Within the framework of CP conservation, the authors of Ref. \cite{heinemeyer1,heinemeyer2}
present the supersymmetric corrections from some special two-loop diagrams
where a close chargino (neutralino) or scalar fermion loop is
inserted into those two-Higgs-doublet one-loop diagrams. Ref. \cite{geng}
discusses the contributions to the muon MDM from the effective
vertices $H^\pm W^\mp\gamma, h_0(H_0)\gamma\gamma$ which are induced
by the scalar quarks of the third generation in the minimal
supersymmetric extension of SM. In the split scenario,
the supersymmetric contributions to the electric dipole moment (EDM)
of lepton have been already presented in
Ref. \cite{DChang,Giudice}. Under the assumption $|m_{1,2}|,|\mu_{_H}|\gg m_{_{\rm w}}$,
\footnote{$m_2,\;m_1$ denote the masses of $SU(2)\times U(1)$ gauginos in the soft breaking
terms, and $\mu_{_H}$ denotes the $\mu$-parameter in the superpotential, respectively.}
the authors derive 5 CP-odd dimension 6 operators involving gauge fields and
Higgs after they integrate out charginos and neutralinos at one-loop level.
Inserting the effective couplings from those CP-odd dimension-6 operators
into those relevant SM one-loop diagrams, they then get the lepton EDMs.

\begin{figure}[t]
\setlength{\unitlength}{1mm}
\begin{center}
\begin{picture}(0,120)(0,0)
\put(-60,-40){\includegraphics{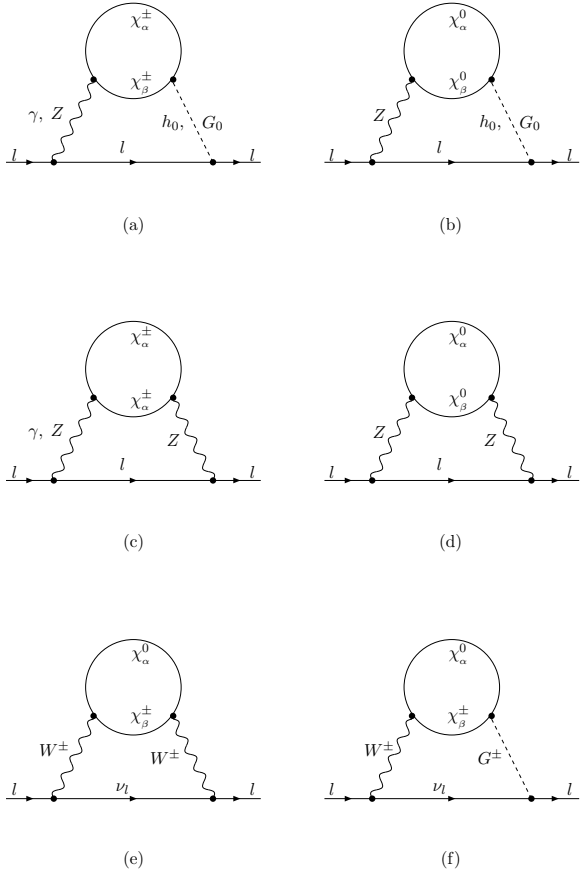}}
\end{picture}
\caption[]{The two-loop self energy diagrams which lead to the
lepton MDMs and EDMs in  split SUSY, the corresponding triangle diagrams
are obtained by attaching a photon in all possible ways
to the internal particles. In concrete calculation, the contributions from
those mirror diagrams should be included also.}
\label{fig1}
\end{center}
\end{figure}

In this paper, we apply the effective Lagrangian method to get the
anomalous dipole moments of lepton. The effective Lagrangian method
has been adopted to calculate the two-loop supersymmetric corrections
to the branching ratio of $b\rightarrow s\gamma$ \cite{Feng1}, neutron EDM \cite{Feng2}
and lepton MDMs and EDMs \cite{Feng3}. In concrete
calculation, we assume that all external leptons as well as photon
are off-shell, then expand the amplitude of corresponding triangle
diagrams according to the external momenta of leptons and photon.
Using loop momentum translational invariance, we formulate the sum of
amplitude from those triangle diagrams which correspond to the corresponding self-energy in
the form which explicitly satisfies the Ward identity required by
the QED gauge symmetry. Then we can get all dimension 6 operators
together with their coefficients. After the equations of
motion are applied to external leptons, higher dimensional operators, such
as dimension 8 operators, also contribute to the muon MDM  and
the electron EDM in principle. However, the contributions of dimension 8
operators contain an additional suppression factor
$m_l^2/\Lambda_{_{\rm NP}}^2$ comparing with that of dimension
6 operators, where $m_l$ is the mass of lepton. Setting
$\Lambda_{_{\rm NP}}\sim100{\rm GeV}$, one finds that this suppression factor is
about $10^{-6}$ for muon, and $10^{-10}$ for electron separatively.
Under current experimental precision, it implies
that the contributions of all higher dimension operators ($D\ge8$)
can be neglected safely.

We adopt the naive dimensional regularization with the
anticommuting $\gamma_{_5}$ scheme, where there is no distinction
between the first 4 dimensions and the remaining $D-4$ dimensions.
Since the bare effective Lagrangian contains the ultraviolet
divergence which is induced by divergent subdiagrams, we give the
renormalized results in the on-mass-shell scheme \cite{onshell}.
Additional, we adopt the nonlinear $R_\xi$ gauge with $\xi=1$ for
simplification \cite{nonlinear-R-xi}. This special gauge-fixing term
guarantees explicit electromagnetic gauge invariance throughout the calculation,
not just at the end because the choice of gauge-fixing term
eliminates the $\gamma W^\pm G^\mp$ vertex in the Lagrangian.

Since the lepton EDM is an interesting topic in both
theoretical and experimental aspects \cite{nexp}, the current experimental upper limit
on the electron EDM is $1.7\times10^{-27}e\cdot cm$ at 95\% CL\cite{Regan},
and a future experiment with precision of $10^{-29}e\cdot cm$ is also proposed\cite{Kawell},
we as well present the lepton EDM by keeping all possible CP violating phases.
Certainly, some diagrams in Fig.\ref{fig1} have been discussed in Ref.\cite{heinemeyer2}
where the authors apply the projecting operator to get the lepton MDMs
(Eq.8$\sim$Eq.10 in Ref.\cite{heinemeyer2}).
Nevertheless, the substantive corrections from several diagrams
are ignored unreasonably (Fig.5 in Ref.\cite{heinemeyer2}).
Additional, our formulae are new in their analytical forms.
For the analysis on the electron EDM, we also include the contributions
from the self energy diagrams (c) and (f) that are neglected
in Ref.\cite{DChang,Giudice}. Our result is more universal than that of Ref.\cite{DChang,Giudice}
because we give up the assumption $|m_{1,2}|,|\mu_{_H}|\gg m_{_{\rm w}}$
in concrete analysis.

This paper is composed by the sections as follows.
In section \ref{sec2}, we introduce the effective Lagrangian method and our notations.
Then we will demonstrate how to obtain the supersymmetric two-loop corrections
to the lepton MDMs and EDMs.
Section \ref{sec3} is devoted to the numerical analysis and
discussion. In section \ref{sec4}, we give our conclusion. Some
tedious formulae are collected in appendix.

\section{Notations and two-loop supersymmetric corrections \label{sec2}}
\indent\indent
The lepton MDMs and EDMs can actually be expressed as the operators
\begin{eqnarray}
&&{\cal L}_{_{MDM}}={e\over4m_{_l}}\;a_{_l}\;\bar{l}\sigma^{\mu\nu}
l\;F_{_{\mu\nu}}
\;,\nonumber\\
&&{\cal L}_{_{EDM}}=-{i\over2}\;d_{_l}\;\bar{l}\sigma^{\mu\nu}\gamma_5
l\;F_{_{\mu\nu}}\;.
\label{adm}
\end{eqnarray}
Here $\sigma_{\mu\nu}=i[\gamma_\mu,\gamma_\nu]/2$, $l$ denotes the lepton fermion,
$F_{_{\mu\nu}}$ is the electromagnetic field strength, $m_{_l}$ is the lepton mass and
$e$ represents the electric charge. Note that the lepton here is on-shell.

In fact, it is convenient to get the corrections from loop diagrams
to lepton MDMs and EDMs in terms of the effective Lagrangian method, if the masses of
internal lines are much heavier than the external lepton mass. Assuming external leptons
as well as photon are all off-shell, we expand the amplitude of the
corresponding triangle diagrams according to the external momenta of
leptons and photon. Then we can get all high dimension operators together
with their coefficients. As discussed in the section \ref{sec1}, it is enough to
retain only those dimension 6 operators in later calculations:
\begin{eqnarray}
&&{\cal O}_{_1}^\mp={1\over(4\pi)^2}\;\bar{l}\;(i/\!\!\!\!{\cal D})^3
\omega_\mp\;l\;,\nonumber\\
&&{\cal O}_{_2}^\mp={e\over(4\pi)^2}\;\overline{(i{\cal D}_{_\mu}l)}
\gamma^\mu F\cdot\sigma\omega_\mp l\;,\nonumber\\
&&{\cal O}_{_3}^\mp={e\over(4\pi)^2}\;\bar{l}F\cdot\sigma\gamma^\mu
\omega_\mp(i{\cal D}_{_\mu}l)\;,\nonumber\\
&&{\cal O}_{_4}^\mp={e\over(4\pi)^2}\;\bar{l}(\partial^\mu F_{_{\mu\nu}})
\gamma^\nu\omega_\mp l\;,\nonumber\\
&&{\cal O}_{_5}^\mp={m_{_l}\over(4\pi)^2}\;\bar{l}\;(i/\!\!\!\!{\cal D})^2
\omega_\mp\;l\;,\nonumber\\
&&{\cal O}_{_6}^\mp={eQ_{_f}m_{_l}\over(4\pi)^2}\;\bar{l}\;F\cdot\sigma
\omega_\mp\;l\;,\nonumber\\
\label{ops}
\end{eqnarray}
with ${\cal D}_{_\mu}=\partial_{_\mu}+ieA_{_\mu}$ and $\omega_\mp=(1\mp\gamma_5)/2$.
When the equations of motion are applied to the incoming and outgoing leptons separately,
only the operators ${\cal O}_{_{2,3,6}}^\mp$ actually contribute to the MDMs and EDMs of leptons.
We will only present the Wilson coefficients of the operators ${\cal O}_{_{2,3,6}}^\mp$
in the effective Lagrangian in our following narration
because of the reason mentioned above. We will adopt below a
terminology where, for example, the "$\gamma h_0$" contribution means the sum of amplitude
from those triangle diagrams (indeed three triangles bound together), in which
a closed fermion (chargino/neutralino) loop is attached to  the virtual Higgs
and photon fields with a real photon attached in all possible ways to the
internal lines. Because the sum of amplitude from those "triangle" diagrams
corresponding to each "self-energy" obviously respects the Ward
identity requested by QED gauge symmetry, we can calculate the
contributions of all the "self-energies" separately. Taking the same steps which
we did in our earlier works \cite{Feng1,Feng2,Feng3}, we obtain the
effective Lagrangian that originates from the self energy diagrams in Fig.\ref{fig1}.
In the bare effective Lagrangian from the 'WW' and 'ZZ' contributions,
the ultraviolet divergence caused by divergent sub-diagrams can be subtracted safely
in on-mass-shell scheme \cite{onshell}. Now, we present the effective Lagrangian
corresponding to the diagrams in Fig.\ref{fig1} respectively.

\subsection{The effective Lagrangian from $\gamma h_0$ ($\gamma G_0$) sector}
\indent\indent
As a closed chargino loop is attached to the virtual neutral Higgs and photon
fields, a real photon can be emitted from either the virtual lepton or the virtual charginos
in the self energy diagram. When a real photon is emitted
from the virtual charginos, the corresponding "triangle" diagrams
belong to the typical two-loop Bar-Zee-type diagrams \cite{Barr-Zee}.
Within the framework of minimal supersymmetric extension of the SM,
the contributions from two-loop Bar-Zee-type diagrams to the EDMs of those light fermions
are discussed extensively in literature \cite{BZ-mssm}.
When a real photon is attached to the internal standard fermion,
the correction from corresponding triangle diagram to
the effective Lagrangian is zero because of the Furry theorem,
this point is also verified through a strict analysis.
The corresponding effective Lagrangian from this sector is written as
\begin{eqnarray}
&&{\cal L}_{_{\gamma h}}=
{e^4\over2\sqrt{2}(4\pi)^2s_{_{\rm w}}^2\Lambda^2}\Bigg\{\Re({\cal H}_{\alpha\alpha})
\Big({x_{_{\chi_\alpha^\pm}}\over x_{_{\rm w}}}\Big)^{1/2}
T_1(x_{_h},x_{_{\chi_\alpha^\pm}},x_{_{\chi_\alpha^\pm}})
\Big({\cal O}_{_6}^++{\cal O}_{_6}^-\Big)
\nonumber\\
&&\hspace{1.2cm}
+i\Im({\cal H}_{\alpha\alpha})\Big({x_{_{\chi_\alpha^\pm}}\over x_{_{\rm w}}}\Big)^{1/2}
T_2(x_{_h},x_{_{\chi_\alpha^\pm}},x_{_{\chi_\alpha^\pm}})
\Big({\cal O}_{_6}^+-{\cal O}_{_6}^-\Big)\Bigg\}
\label{gamma-h}
\end{eqnarray}
with
\begin{eqnarray}
&&{\cal H}_{\alpha\beta}=(U_{_R}^\dagger)_{_{\alpha2}}(U_{_L})_{_{1\beta}}\cos\beta
+(U_{_R}^\dagger)_{_{\alpha1}}(U_{_L})_{_{2\beta}}\sin\beta\;.
\label{coupling-h}
\end{eqnarray}
Where $U_{_{L,R}}$ denote the left- and right-mixing matrices of charginos,
$\Lambda$ denotes a energy scale to define $x_i=m_i^2/\Lambda^2$, respectively.
The angle $\beta$ is defined through the ratio between the vacuum expectation
values of two Higgs doublets: $\tan\beta=\upsilon_2/\upsilon_1$. We adopt the shortcut
notations: $c_{_{\rm w}}=\cos\theta_{_{\rm w}},\;s_{_{\rm w}}
=\sin\theta_{_{\rm w}},\;$ where $\theta_{_{\rm w}}$ is the Weinberg angle.
The concrete expressions of $T_{1,2}$ can be found in appendix.

Accordingly, the lepton MDMs and EDMs from $\gamma h_0$ sector are written as
\begin{eqnarray}
&&a_l^{\gamma h}={\sqrt{2}e^4Q_{_f}m_{_l}^2\over(4\pi)^4s_{_{\rm w}}^2\Lambda^2}
\Re({\cal H}_{\alpha\alpha})\Big({x_{_{\chi_\alpha^\pm}}\over x_{_{\rm w}}}\Big)^{1/2}
T_1(x_{_h},x_{_{\chi_\alpha^\pm}},x_{_{\chi_\alpha^\pm}})\;,
\nonumber\\
&&d_l^{\gamma h}=-{e^5Q_{_f}m_{_l}\over\sqrt{2}(4\pi)^4s_{_{\rm w}}^2\Lambda^2}
\Im({\cal H}_{\alpha\alpha})\Big({x_{_{\chi_\alpha^\pm}}\over x_{_{\rm w}}}\Big)^{1/2}
T_2(x_{_h},x_{_{\chi_\alpha^\pm}},x_{_{\chi_\alpha^\pm}})\;.
\label{MD-gamma-h}
\end{eqnarray}
In the limit $x_{_{\chi_\alpha^\pm}}\gg x_{_h}$, the above expressions can
be simplified as
\begin{eqnarray}
&&a_l^{\gamma h}=-{\sqrt{2}e^4Q_{_f}m_{_l}^2\over(4\pi)^4s_{_{\rm w}}^2\Lambda^2}
\Re({\cal H}_{\alpha\alpha})\Big({x_{_{\chi_\alpha^\pm}}\over x_{_{\rm w}}}\Big)^{1/2}
\lim\limits_{x_{_{\chi_\beta^\pm}}\rightarrow x_{_{\chi_\alpha^\pm}}}
{\partial\over\partial x_{_{\chi_\beta^\pm}}}\varphi_1(x_{_{\chi_\alpha^\pm}},
x_{_{\chi_\beta^\pm}})\;,
\nonumber\\
&&d_l^{\gamma h}=-{e^5Q_{_f}m_{_l}\over\sqrt{2}(4\pi)^4s_{_{\rm w}}^2\Lambda^2}
\Im({\cal H}_{\alpha\alpha})\Big({x_{_{\chi_\alpha^\pm}}\over x_{_{\rm w}}}\Big)^{1/2}
\Big[{\ln x_{_h}\over x_{_{\chi_\alpha^\pm}}}
+\lim\limits_{x_{_{\chi_\beta^\pm}}\rightarrow x_{_{\chi_\alpha^\pm}}}
{\partial\over\partial x_{_{\chi_\beta^\pm}}}\varphi_1(x_{_{\chi_\alpha^\pm}},
x_{_{\chi_\beta^\pm}})\Big]\;.
\label{MD-gamma-h1}
\end{eqnarray}

Similarly, we can formulate the corrections from $\gamma G_0$ sector to
the effective Lagrangian as
\begin{eqnarray}
&&{\cal L}_{_{\gamma G}}=
{e^4\over2\sqrt{2}(4\pi)^2s_{_{\rm w}}^2\Lambda^2}\Bigg\{\Re({\cal H}_{\alpha\alpha})
\Big({x_{_{\chi_\alpha^\pm}}\over x_{_{\rm w}}}\Big)^{1/2}
T_2(x_{_{\rm z}},x_{_{\chi_\alpha^\pm}},x_{_{\chi_\beta^\pm}})
\Bigg]\Big({\cal O}_{_6}^++{\cal O}_{_6}^-\Big)
\nonumber\\
&&\hspace{1.2cm}
-i\Im({\cal H}_{\alpha\alpha})\Big({x_{_{\chi_\alpha^\pm}}\over x_{_{\rm w}}}\Big)^{1/2}
T_1(x_{_{\rm z}},x_{_{\chi_\alpha^\pm}},x_{_{\chi_\beta^\pm}})
\Big({\cal O}_{_6}^+-{\cal O}_{_6}^-\Big)\Bigg\}\;.
\label{gamma-G}
\end{eqnarray}
Correspondingly, the corrections to the lepton MDMs and EDMs from this
sector are:
\begin{eqnarray}
&&a_l^{\gamma G}={\sqrt{2}e^4Q_{_f}m_{_l}^2\over(4\pi)^4s_{_{\rm w}}^2\Lambda^2}
\Re({\cal H}_{\alpha\alpha})\Big({x_{_{\chi_\alpha^\pm}}\over x_{_{\rm w}}}\Big)^{1/2}
T_2(x_{_{\rm z}},x_{_{\chi_\alpha^\pm}},x_{_{\chi_\alpha^\pm}})\;,
\nonumber\\
&&d_l^{\gamma G}={e^5Q_{_f}m_{_l}\over\sqrt{2}(4\pi)^4s_{_{\rm w}}^2\Lambda^2}
\Im({\cal H}_{\alpha\alpha})\Big({x_{_{\chi_\alpha^\pm}}\over x_{_{\rm w}}}\Big)^{1/2}
T_1(x_{_{\rm z}},x_{_{\chi_\alpha^\pm}},x_{_{\chi_\alpha^\pm}})\;.
\label{MD-gamma-G}
\end{eqnarray}
In the limit $x_{_{\chi_\alpha^\pm}}\gg x_{_{\rm z}}$, we have
\begin{eqnarray}
&&a_l^{\gamma G}={\sqrt{2}e^4Q_{_f}m_{_l}^2\over(4\pi)^4s_{_{\rm w}}^2\Lambda^2}
\Re({\cal H}_{\alpha\alpha})\Big({x_{_{\chi_\alpha^\pm}}\over x_{_{\rm w}}}\Big)^{1/2}
\Big[{\ln x_{_{\rm z}}\over x_{_{\chi_\alpha^\pm}}}
+\lim\limits_{x_{_{\chi_\beta^\pm}}\rightarrow x_{_{\chi_\alpha^\pm}}}
{\partial\over\partial x_{_{\chi_\beta^\pm}}}\varphi_1(x_{_{\chi_\alpha^\pm}},
x_{_{\chi_\beta^\pm}})\Big]\;,
\nonumber\\
&&d_l^{\gamma G}={e^5Q_{_f}m_{_l}\over\sqrt{2}(4\pi)^4s_{_{\rm w}}^2\Lambda^2}
\Im({\cal H}_{\alpha\alpha})\Big({x_{_{\chi_\alpha^\pm}}\over x_{_{\rm w}}}\Big)^{1/2}
\lim\limits_{x_{_{\chi_\beta^\pm}}\rightarrow x_{_{\chi_\alpha^\pm}}}
{\partial\over\partial x_{_{\chi_\beta^\pm}}}\varphi_1(x_{_{\chi_\alpha^\pm}},
x_{_{\chi_\beta^\pm}})\;.
\label{MD-gamma-G1}
\end{eqnarray}

It should be emphasized that the corrections from this sector
to the lepton EDMs are neglected in the analysis before \cite{DChang,Giudice}.
However, Eq.\ref{MD-gamma-G} implies that the contributions from those diagrams to the
lepton MDMs and EDMs can not be ignored generally.

Using the concrete expression of $\varphi_1(x,y)$ collected in appendix,
one can verify easily that the corrections to the lepton MDMs and EDMs
from the sectors are suppressed by the masses of charginos as
$m_{_{\chi_\alpha^\pm}}\gg m_{_h},\;m_{_{\rm z}}\;(\alpha=1,\;2)$.

\subsection{The effective Lagrangian from $Zh_0$ ($ZG_0$) sector}
\indent\indent
As a closed chargino loop is attached to the virtual Higgs and $Z$ gauge boson
fields, a real photon can be attached to either the virtual lepton or the virtual charginos
in the self energy diagram. When a real photon is attached to the virtual lepton,
the corresponding amplitude only modifies the Wilson coefficients of the
operators ${\cal O}_{_5}^\pm$ in the effective Lagrangian after the heavy freedoms
are integrated out. In other words, this triangle diagram does not contribute to the
lepton MDMs and EDMs. A real photon can be only attached to the virtual lepton
as the closed loop is composed of neutralinos,
the corresponding triangle diagram does not affect the theoretical predictions
on the lepton MDMs and EDMs for the same reason.
Considering the points above, we formulate the contributions from $Zh_0$ sector
to the effective Lagrangian as
\begin{eqnarray}
&&{\cal L}_{_{Zh}}=
-{e^4\over16\sqrt{2}(4\pi)^2s_{_{\rm w}}^4c_{_{\rm w}}^2Q_{_f}\Lambda^2}
(T_{_f}^Z-2Q_{_f}s_{_{\rm w}}^2)\Bigg\{\Big({x_{_{\chi_\beta^\pm}}\over x_{_{\rm w}}}\Big)^{1/2}
\Big[2(2+\ln x_{_{\chi_\beta^\pm}})\varrho_{_{0,1}}(x_{_{\rm z}},x_{_h})
\nonumber\\
&&\hspace{1.2cm}
+F_1(x_{_{\rm z}},x_{_h},x_{_{\chi_\alpha^\pm}},x_{_{\chi_\beta^\pm}})\Big]
\Re\Big({\cal H}_{_{\beta\alpha}}\xi^L_{_{\alpha\beta}}+{\cal H}^\dagger_{_{\beta\alpha}}
\xi^R_{_{\alpha\beta}}\Big)({\cal O}_{_6}^++{\cal O}_{_6}^-)
\nonumber\\
&&\hspace{1.2cm}
+i\Big({x_{_{\chi_\beta^\pm}}\over x_{_{\rm w}}}\Big)^{1/2}
\Big[-2(\ln x_{_{\chi_\alpha^\pm}}-\ln x_{_{\chi_\beta^\pm}})
\varrho_{_{0,1}}(x_{_{\rm z}},x_{_h})
+F_1(x_{_{\rm z}},x_{_h},x_{_{\chi_\alpha^\pm}},x_{_{\chi_\beta^\pm}})
\nonumber\\
&&\hspace{1.2cm}
+F_2(x_{_{\rm z}},x_{_h},x_{_{\chi_\beta^\pm}},x_{_{\chi_\alpha^\pm}})\Big]
\Im\Big({\cal H}_{_{\beta\alpha}}\xi^L_{_{\alpha\beta}}-{\cal H}^\dagger_{_{\beta\alpha}}
\xi^R_{_{\alpha\beta}}\Big)({\cal O}_{_6}^--{\cal O}_{_6}^+)\Bigg\}+\cdots
\label{zh}
\end{eqnarray}
with
\begin{eqnarray}
&&\xi^L_{\alpha\beta}=2\delta_{\alpha\beta}\cos2\theta_{_{\rm w}}
+(U_{_L}^\dagger)_{_{\alpha1}}(U_{_L})_{_{1\beta}}
\;,\nonumber\\
&&\xi^R_{\alpha\beta}=2\delta_{\alpha\beta}\cos2\theta_{_{\rm w}}
+(U_{_R}^\dagger)_{_{\alpha1}}(U_{_R})_{_{1\beta}}\;,
\label{coupling-xi}
\end{eqnarray}
where the concrete expressions of the functions $\varrho_{_{i,j}}(x_1,x_2),\;
F_{1,2}(x_1,x_2,x_3,x_4)$ are listed in appendix. Additional, $T_{_f}^Z$
is the isospin of lepton, and $Q_{_f}$ is the electric charge of lepton, respectively.
Using Eq.\ref{zh}, we get the corrections to the lepton MDMs and EDMs from $Zh_0$ sector as
\begin{eqnarray}
&&a_l^{Zh}=-{e^4m_{_l}^2\over4\sqrt{2}(4\pi)^4s_{_{\rm w}}^4c_{_{\rm w}}^2\Lambda^2}
(T_{_f}^Z-2Q_{_f}s_{_{\rm w}}^2)\Big({x_{_{\chi_\beta^\pm}}\over x_{_{\rm w}}}\Big)^{1/2}
\Big[2(2+\ln x_{_{\chi_\beta^\pm}})\varrho_{_{i,j}}(x_{_{\rm z}},x_{_h})
\nonumber\\
&&\hspace{1.2cm}
+F_1(x_{_{\rm z}},x_{_h},x_{_{\chi_\alpha^\pm}},x_{_{\chi_\beta^\pm}})\Big]
\Re\Big({\cal H}_{_{\beta\alpha}}\xi^L_{_{\alpha\beta}}+{\cal H}^\dagger_{_{\beta\alpha}}
\xi^R_{_{\alpha\beta}}\Big)\;,
\nonumber\\
&&d_l^{Zh}={e^5m_{_l}\over8\sqrt{2}(4\pi)^4s_{_{\rm w}}^4c_{_{\rm w}}^2\Lambda^2}
(T_{_f}^Z-2Q_{_f}s_{_{\rm w}}^2)\Big({x_{_{\chi_\beta^\pm}}\over x_{_{\rm w}}}\Big)^{1/2}
\Big[-2(\ln x_{_{\chi_\alpha^\pm}}-\ln x_{_{\chi_\beta^\pm}})
\varrho_{_{0,1}}(x_{_{\rm z}},x_{_h})
\nonumber\\
&&\hspace{1.2cm}
+F_1(x_{_{\rm z}},x_{_h},x_{_{\chi_\alpha^\pm}},x_{_{\chi_\beta^\pm}})
+F_2(x_{_{\rm z}},x_{_h},x_{_{\chi_\beta^\pm}},x_{_{\chi_\alpha^\pm}})\Big]
\Im\Big({\cal H}_{_{\beta\alpha}}\xi^L_{_{\alpha\beta}}-{\cal H}^\dagger_{_{\beta\alpha}}
\xi^R_{_{\alpha\beta}}\Big)\;.
\label{MD-z-h}
\end{eqnarray}
The above equations contain the suppression factor $1-4s_{_{\rm w}}^2$ because $Q_{_f}=-1$
and $T_{_f}^Z=-1/2$ for charged leptons. In the limit $x_{_{\chi_\alpha^\pm}},\;
x_{_{\chi_\beta^\pm}}\gg x_{_{\rm z}},\;x_{_h}$, Eq.\ref{MD-z-h} can be approximated as
\begin{eqnarray}
&&a_l^{Zh}=-{e^4m_{_l}^2\over4\sqrt{2}(4\pi)^4s_{_{\rm w}}^4c_{_{\rm w}}^2\Lambda^2}
(T_{_f}^Z-2Q_{_f}s_{_{\rm w}}^2)\Big({x_{_{\chi_\beta^\pm}}\over x_{_{\rm w}}}\Big)^{1/2}
\Big[{\partial\varphi_1\over\partial x_{_{\chi_\beta^\pm}}}(x_{_{\chi_\alpha^\pm}}
,x_{_{\chi_\beta^\pm}})
\nonumber\\
&&\hspace{1.2cm}
-{2-2x_{_{\chi_\alpha^\pm}}\varrho_{_{0,1}}(x_{_{\chi_\alpha^\pm}},x_{_{\chi_\beta^\pm}})
\over x_{_{\chi_\alpha^\pm}}-x_{_{\chi_\beta^\pm}}}\cdot
\varrho_{_{1,1}}(x_{_{\rm z}},x_{_h})\Big]
\Re\Big({\cal H}_{_{\beta\alpha}}\xi^L_{_{\alpha\beta}}+{\cal H}^\dagger_{_{\beta\alpha}}
\xi^R_{_{\alpha\beta}}\Big)\;,
\nonumber\\
&&d_l^{Zh}={e^5m_{_l}\over8\sqrt{2}(4\pi)^4s_{_{\rm w}}^4c_{_{\rm w}}^2\Lambda^2}
(T_{_f}^Z-2Q_{_f}s_{_{\rm w}}^2)\Big({x_{_{\chi_\beta^\pm}}\over x_{_{\rm w}}}\Big)^{1/2}
\Big[\Big({\partial\varphi_1\over\partial x_{_{\chi_\alpha^\pm}}}
+{\partial\varphi_1\over\partial x_{_{\chi_\beta^\pm}}}\Big)
(x_{_{\chi_\alpha^\pm}},x_{_{\chi_\beta^\pm}})
\nonumber\\
&&\hspace{1.2cm}
+2\varrho_{_{0,1}}(x_{_{\chi_\alpha^\pm}},x_{_{\chi_\beta^\pm}})
\varrho_{_{1,1}}(x_{_{\rm z}},x_{_h})\Big]
\Im\Big({\cal H}_{_{\beta\alpha}}\xi^L_{_{\alpha\beta}}-{\cal H}^\dagger_{_{\beta\alpha}}
\xi^R_{_{\alpha\beta}}\Big)\;.
\label{MD-z-h1}
\end{eqnarray}

Similarly, the contributions from $ZG_0$ sector to the effective Lagrangian are
\begin{eqnarray}
&&{\cal L}_{_{ZG_0}}=
-{e^4\over16\sqrt{2}(4\pi)^2s_{_{\rm w}}^4c_{_{\rm w}}^2Q_{_f}\Lambda^2}
\Bigg\{-i\Big({x_{_{\chi_\beta^\pm}}\over x_{_{\rm w}}}\Big)^{1/2}
\Big[{2\over x_{_{\rm z}}}(2+\ln x_{_{\chi_\beta^\pm}})
+F_1(x_{_{\rm z}}, x_{_{\rm z}},x_{_{\chi_\alpha^\pm}},x_{_{\chi_\beta^\pm}})\Big]
\nonumber\\
&&\hspace{1.2cm}\times
\Im\Big({\cal H}_{_{\beta\alpha}}\xi^L_{_{\alpha\beta}}+{\cal H}^\dagger_{_{\beta\alpha}}
\xi^R_{_{\alpha\beta}}\Big)
(T_{_f}^Z-2Q_{_f}s_{_{\rm w}}^2)({\cal O}_{_6}^--{\cal O}_{_6}^+)
\nonumber\\
&&\hspace{1.2cm}
+\Big({x_{_{\chi_\beta^\pm}}\over x_{_{\rm w}}}\Big)^{1/2}
\Big[-{2\over  x_{_{\rm z}}}(\ln x_{_{\chi_\alpha^\pm}}-\ln x_{_{\chi_\beta^\pm}})
+F_1(x_{_{\rm z}},x_{_{\rm z}},x_{_{\chi_\alpha^\pm}},x_{_{\chi_\beta^\pm}})
+F_2(x_{_{\rm z}},x_{_{\rm z}},x_{_{\chi_\beta^\pm}},x_{_{\chi_\alpha^\pm}})\Big]
\nonumber\\
&&\hspace{1.2cm}\times
\Re\Big({\cal H}_{_{\beta\alpha}}\xi^L_{_{\alpha\beta}}-{\cal H}^\dagger_{_{\beta\alpha}}
\xi^R_{_{\alpha\beta}}\Big)(T_{_f}^Z-2Q_{_f}s_{_{\rm w}}^2)({\cal O}_{_6}^-+{\cal O}_{_6}^+)\Bigg\}
+\cdots\;,
\label{zG}
\end{eqnarray}
and the contributions to the lepton MDMs and EDMs are:
\begin{eqnarray}
&&a_l^{ZG}=-{e^4m_{_l}^2\over4\sqrt{2}(4\pi)^4s_{_{\rm w}}^4c_{_{\rm w}}^2\Lambda^2}
(T_{_f}^Z-2Q_{_f}s_{_{\rm w}}^2)\Big({x_{_{\chi_\beta^\pm}}\over x_{_{\rm w}}}\Big)^{1/2}
\Big[-{2\over  x_{_{\rm z}}}(\ln x_{_{\chi_\alpha^\pm}}-\ln x_{_{\chi_\beta^\pm}})
\nonumber\\
&&\hspace{1.2cm}
+F_1(x_{_{\rm z}},x_{_{\rm z}},x_{_{\chi_\alpha^\pm}},x_{_{\chi_\beta^\pm}})
+F_2(x_{_{\rm z}},x_{_{\rm z}},x_{_{\chi_\beta^\pm}},x_{_{\chi_\alpha^\pm}})\Big]
\Re\Big({\cal H}_{_{\beta\alpha}}\xi^L_{_{\alpha\beta}}-{\cal H}^\dagger_{_{\beta\alpha}}
\xi^R_{_{\alpha\beta}}\Big)\;,
\nonumber\\
&&d_l^{ZG}={e^5m_{_l}\over8\sqrt{2}(4\pi)^4s_{_{\rm w}}^4c_{_{\rm w}}^2\Lambda^2}
(T_{_f}^Z-2Q_{_f}s_{_{\rm w}}^2)\Big({x_{_{\chi_\beta^\pm}}\over x_{_{\rm w}}}\Big)^{1/2}
\Big[{2\over x_{_{\rm z}}}(2+\ln x_{_{\chi_\beta^\pm}})
\nonumber\\
&&\hspace{1.2cm}
+F_1(x_{_{\rm z}}, x_{_{\rm z}},x_{_{\chi_\alpha^\pm}},x_{_{\chi_\beta^\pm}})\Big]
\Im\Big({\cal H}_{_{\beta\alpha}}\xi^L_{_{\alpha\beta}}+{\cal H}^\dagger_{_{\beta\alpha}}
\xi^R_{_{\alpha\beta}}\Big)\;.
\label{MD-z-G}
\end{eqnarray}
When $x_{_{\chi_\alpha^\pm}},\;x_{_{\chi_\beta^\pm}}\gg x_{_{\rm z}}$,
Eq.\ref{MD-z-G} can be approached by
\begin{eqnarray}
&&a_l^{ZG}=-{e^4m_{_l}^2\over4\sqrt{2}(4\pi)^4s_{_{\rm w}}^4c_{_{\rm w}}^2\Lambda^2}
(T_{_f}^Z-2Q_{_f}s_{_{\rm w}}^2)\Big({x_{_{\chi_\beta^\pm}}\over x_{_{\rm w}}}\Big)^{1/2}
\Big[\Big({\partial\varphi_1\over\partial x_{_{\chi_\alpha^\pm}}}
+{\partial\varphi_1\over\partial x_{_{\chi_\beta^\pm}}}\Big)
(x_{_{\chi_\alpha^\pm}},x_{_{\chi_\beta^\pm}})
\nonumber\\
&&\hspace{1.2cm}
+2(1+\ln x_{_{\rm z}})\varrho_{_{0,1}}(x_{_{\chi_\alpha^\pm}},x_{_{\chi_\beta^\pm}})\Big]
\Re\Big({\cal H}_{_{\beta\alpha}}\xi^L_{_{\alpha\beta}}-{\cal H}^\dagger_{_{\beta\alpha}}
\xi^R_{_{\alpha\beta}}\Big)\;,
\nonumber\\
&&d_l^{ZG}={e^5m_{_l}\over8\sqrt{2}(4\pi)^4s_{_{\rm w}}^4c_{_{\rm w}}^2\Lambda^2}
(T_{_f}^Z-2Q_{_f}s_{_{\rm w}}^2)\Big({x_{_{\chi_\beta^\pm}}\over x_{_{\rm w}}}\Big)^{1/2}
\Big[{\partial\varphi_1\over\partial x_{_{\chi_\beta^\pm}}}(x_{_{\chi_\alpha^\pm}}
,x_{_{\chi_\beta^\pm}})
\nonumber\\
&&\hspace{1.2cm}
-{2-2x_{_{\chi_\alpha^\pm}}\varrho_{_{0,1}}(x_{_{\chi_\alpha^\pm}},x_{_{\chi_\beta^\pm}})
\over x_{_{\chi_\alpha^\pm}}-x_{_{\chi_\beta^\pm}}}\cdot(1+\ln x_{_{\rm z}})\Big]
\Im\Big({\cal H}_{_{\beta\alpha}}\xi^L_{_{\alpha\beta}}+{\cal H}^\dagger_{_{\beta\alpha}}
\xi^R_{_{\alpha\beta}}\Big)\;.
\label{MD-z-G1}
\end{eqnarray}

\subsection{The effective Lagrangian from $\gamma Z$ sector}
\indent\indent
When a closed chargino loop is attached to the virtual $\gamma$ and $Z$ gauge bosons,
the corresponding correction to the effective Lagrangian is very tedious. If we ignore the terms
which are proportional to the suppression factor $1-4s_{_{\rm w}}^2$, the correction
from this sector to the effective Lagrangian is drastically simplified as
\begin{eqnarray}
&&{\cal L}_{_{\gamma Z}}=
{e^4\over8(4\pi)^2s_{_{\rm w}}^2c_{_{\rm w}}^2\Lambda^2}
\Big(\xi^L_{_{\alpha\alpha}}-\xi^R_{_{\alpha\alpha}}\Big)
\lim\limits_{x_{_{\chi_\alpha^\pm}}\rightarrow x_{_{\chi_\beta^\pm}}}
T_3(x_{_{\rm z}},x_{_{\chi_\alpha^\pm}},x_{_{\chi_\beta^\pm}})
\nonumber\\
&&\hspace{1.2cm}\times
\Big[\Big(T_{_f}^Z-Q_{_f}s_{_{\rm w}}^2\Big)({\cal O}_{_2}^-+{\cal O}_{_3}^-)
+Q_{_f}s_{_{\rm w}}^2({\cal O}_{_2}^++{\cal O}_{_3}^+)\Big]+\cdots\;.
\label{gamma-z}
\end{eqnarray}
Correspondingly, the correction to the lepton MDMs from this sector is written as
\begin{eqnarray}
&&a_l^{\gamma Z}={e^4m_{_l}^2\over4(4\pi)^4s_{_{\rm w}}^2c_{_{\rm w}}^2\Lambda^2}
\Big(\xi^L_{_{\alpha\alpha}}-\xi^R_{_{\alpha\alpha}}\Big)
\lim\limits_{x_{_{\chi_\beta^\pm}}\rightarrow x_{_{\chi_\alpha^\pm}}}
T_3(x_{_{\rm z}},x_{_{\chi_\alpha^\pm}},x_{_{\chi_\beta^\pm}})\;,
\label{MD-gamma-z}
\end{eqnarray}
and the correction to the lepton EDMs is zero. In the limit
$x_{_{\chi_\alpha^\pm}}\gg x_{_{\rm z}}$, we can approximate the correction
to the lepton MDMs from this sector as
\begin{eqnarray}
&&a_l^{\gamma Z}={e^4m_{_l}^2\over4(4\pi)^4s_{_{\rm w}}^2c_{_{\rm w}}^2\Lambda^2}
\Big(\xi^L_{_{\alpha\alpha}}-\xi^R_{_{\alpha\alpha}}\Big)\Big[
{13\over18x_{_{\chi_\alpha^\pm}}}+{\ln x_{_{\chi_\alpha^\pm}}-2\ln x_{_{\rm z}}
\over3x_{_{\chi_\alpha^\pm}}}
\nonumber\\
&&\hspace{1.2cm}
+\lim\limits_{x_{_{\chi_\beta^\pm}}\rightarrow x_{_{\chi_\alpha^\pm}}}
\Big(2x_{_{\chi_\alpha^\pm}}{\partial^2\varphi_1\over\partial x_{_{\chi_\alpha^\pm}}^2}
-{\partial\varphi_1\over\partial x_{_{\chi_\alpha^\pm}}}\Big)
(x_{_{\chi_\alpha^\pm}},x_{_{\chi_\beta^\pm}})\Big]\;.
\label{MD-gamma-z1}
\end{eqnarray}

\subsection{The effective Lagrangian from $WG^\pm$ sector}
\indent\indent
As a closed chargino-neutralino loop is attached to the virtual
$W^\pm$ gauge boson and charged goldstone $G^\mp$, the induced
Lagrangian can be written as
\begin{eqnarray}
&&{\cal L}_{_{WG}}=
{e^4\over 16(4\pi)^2s_{_{\rm w}}^4c_{_{\rm w}}Q_{_f}\Lambda^2}
\Bigg\{\Big({x_{_{\chi_\beta^\pm}}\over x_{_{\rm w}}}\Big)^{1/2}
F_3(x_{_{\rm w}},x_{_{\rm w}},x_{_{\chi_\alpha^0}},x_{_{\chi_\beta^\pm}})
\Big[\Big({\cal G}^L_{_{\beta\alpha}}\zeta^L_{_{\alpha\beta}}
+{\cal G}^R_{_{\beta\alpha}}\zeta^R_{_{\alpha\beta}}\Big){\cal O}_{_6}^-
\nonumber\\
&&\hspace{1.2cm}
+\Big(({\cal G}^L)^\dagger_{_{\alpha\beta}}(\zeta^L)^\dagger_{_{\beta\alpha}}
+({\cal G}^R)^\dagger_{_{\alpha\beta}}(\zeta^R)^\dagger_{_{\beta\alpha}}\Big){\cal O}_{_6}^+\Big]
\nonumber\\
&&\hspace{1.2cm}
+\Big({x_{_{\chi_\alpha^0}}\over x_{_{\rm w}}}\Big)^{1/2}
F_4(x_{_{\rm w}},x_{_{\rm w}},x_{_{\chi_\alpha^0}},x_{_{\chi_\beta^\pm}})
\Big[\Big({\cal G}^L_{_{\beta\alpha}}\zeta^R_{_{\alpha\beta}}
+{\cal G}^R_{_{\beta\alpha}}\zeta^L_{_{\alpha\beta}}\Big){\cal O}_{_6}^-
+\Big(({\cal G}^R)^\dagger_{_{\alpha\beta}}(\zeta^L)^\dagger_{_{\beta\alpha}}
\nonumber\\
&&\hspace{1.2cm}
+({\cal G}^L)^\dagger_{_{\alpha\beta}}(\zeta^R)^\dagger_{_{\beta\alpha}}\Big){\cal O}_{_6}^+\Big]
\nonumber\\
&&\hspace{1.2cm}
+\Big({x_{_{\chi_\beta^\pm}}\over x_{_{\rm w}}}\Big)^{1/2}
F_5(x_{_{\rm w}},x_{_{\rm w}},x_{_{\chi_\alpha^0}},x_{_{\chi_\beta^\pm}})
\Big[\Big({\cal G}^L_{_{\beta\alpha}}\zeta^L_{_{\alpha\beta}}
-{\cal G}^R_{_{\beta\alpha}}\zeta^R_{_{\alpha\beta}}\Big){\cal O}_{_6}^-
+\Big(({\cal G}^L)^\dagger_{_{\alpha\beta}}(\zeta^L)^\dagger_{_{\beta\alpha}}
\nonumber\\
&&\hspace{1.2cm}
-({\cal G}^R)^\dagger_{_{\alpha\beta}}(\zeta^R)^\dagger_{_{\beta\alpha}}\Big){\cal O}_{_6}^+\Big]
\nonumber\\
&&\hspace{1.2cm}
+\Big({x_{_{\chi_\alpha^0}}\over x_{_{\rm w}}}\Big)^{1/2}
F_6(x_{_{\rm w}},x_{_{\rm w}},x_{_{\chi_\alpha^0}},x_{_{\chi_\beta^\pm}})
\Big[\Big({\cal G}^L_{_{\beta\alpha}}\zeta^R_{_{\alpha\beta}}
-{\cal G}^R_{_{\beta\alpha}}\zeta^L_{_{\alpha\beta}}\Big){\cal O}_{_6}^-
+\Big(({\cal G}^L)^\dagger_{_{\alpha\beta}}(\zeta^R)^\dagger_{_{\beta\alpha}}
\nonumber\\
&&\hspace{1.2cm}
-({\cal G}^R)^\dagger_{_{\alpha\beta}}(\zeta^L)^\dagger_{_{\beta\alpha}}\Big){\cal O}_{_6}^+\Big]\Bigg\}
\label{wg}
\end{eqnarray}
with
\begin{eqnarray}
&&\zeta^L_{\alpha\beta}={\cal N}^\dagger_{\alpha2}(U_{_R})_{_{1\beta}}
-{1\over\sqrt{2}}{\cal N}^\dagger_{\alpha4}(U_{_R})_{_{2\beta}}
\;,\nonumber\\
&&\zeta^R_{\alpha\beta}={\cal N}_{2\alpha}(U_{_R}^\dagger)_{_{\beta1}}
+{1\over\sqrt{2}}{\cal N}_{3\alpha}(U_{_R}^\dagger)_{_{\beta2}}
\;,\nonumber\\
&&{\cal G}^L_{_{\beta\alpha}}=\sin\beta\Big\{{1\over\sqrt{2}}(U_{_L})_{_{2\beta}}
\Big({\cal N}_{1\alpha}s_{_{\rm w}}+{\cal N}_{2\alpha}c_{_{\rm w}}\Big)
-(U_{_L})_{_{1\beta}}{\cal N}_{3\alpha}c_{_{\rm w}}\Big\}
\;,\nonumber\\
&&{\cal G}^R_{_{\beta\alpha}}=-\cos\beta\Big\{{1\over\sqrt{2}}(U_{_R}^\dagger)_{_{\beta2}}
\Big({\cal N}^\dagger_{\alpha1}s_{_{\rm w}}+{\cal N}^\dagger_{\alpha2}c_{_{\rm w}}\Big)
-(U_{_R}^\dagger)_{_{\beta1}}{\cal N}^\dagger_{\alpha4}c_{_{\rm w}}\Big\}\;.
\label{coupling-zeta}
\end{eqnarray}
Here, the $4\times4$ matrix ${\cal N}$ denotes the mixing matrix of
the four neutralinos $\chi_\alpha^0\;(\alpha=1,\;2,\;3,\;4)$.

The corresponding corrections to the lepton MDMs and EDMs are respectively expressed as
\begin{eqnarray}
&&a_l^{WG}={e^4m_{_l}^2\over4(4\pi)^4s_{_{\rm w}}^4c_{_{\rm w}}\Lambda^2}
\Bigg\{\Big({x_{_{\chi_\beta^\pm}}\over x_{_{\rm w}}}\Big)^{1/2}
F_3(x_{_{\rm w}},x_{_{\rm w}},x_{_{\chi_\alpha^0}},x_{_{\chi_\beta^\pm}})
\Re\Big({\cal G}^L_{_{\beta\alpha}}\zeta^L_{_{\alpha\beta}}
+{\cal G}^R_{_{\beta\alpha}}\zeta^R_{_{\alpha\beta}}\Big)
\nonumber\\
&&\hspace{1.2cm}
+\Big({x_{_{\chi_\alpha^0}}\over x_{_{\rm w}}}\Big)^{1/2}
F_4(x_{_{\rm w}},x_{_{\rm w}},x_{_{\chi_\alpha^0}},x_{_{\chi_\beta^\pm}})
\Re\Big({\cal G}^L_{_{\beta\alpha}}\zeta^R_{_{\alpha\beta}}
+{\cal G}^R_{_{\beta\alpha}}\zeta^L_{_{\alpha\beta}}\Big)
\nonumber\\
&&\hspace{1.2cm}
+\Big({x_{_{\chi_\beta^\pm}}\over x_{_{\rm w}}}\Big)^{1/2}
F_5(x_{_{\rm w}},x_{_{\rm w}},x_{_{\chi_\alpha^0}},x_{_{\chi_\beta^\pm}})
\Re\Big({\cal G}^L_{_{\beta\alpha}}\zeta^L_{_{\alpha\beta}}
-{\cal G}^R_{_{\beta\alpha}}\zeta^R_{_{\alpha\beta}}\Big)
\nonumber\\
&&\hspace{1.2cm}
+\Big({x_{_{\chi_\alpha^0}}\over x_{_{\rm w}}}\Big)^{1/2}
F_6(x_{_{\rm w}},x_{_{\rm w}},x_{_{\chi_\alpha^0}},x_{_{\chi_\beta^\pm}})
\Re\Big({\cal G}^L_{_{\beta\alpha}}\zeta^R_{_{\alpha\beta}}
-{\cal G}^R_{_{\beta\alpha}}\zeta^L_{_{\alpha\beta}}\Big)\Bigg\}\;,
\nonumber\\
&&d_l^{WG}={e^5m_{_l}\over8(4\pi)^4s_{_{\rm w}}^4c_{_{\rm w}}\Lambda^2}
\Bigg\{\Big({x_{_{\chi_\beta^\pm}}\over x_{_{\rm w}}}\Big)^{1/2}
F_3(x_{_{\rm w}},x_{_{\rm w}},x_{_{\chi_\alpha^0}},x_{_{\chi_\beta^\pm}})
\Im\Big({\cal G}^L_{_{\beta\alpha}}\zeta^L_{_{\alpha\beta}}
+{\cal G}^R_{_{\beta\alpha}}\zeta^R_{_{\alpha\beta}}\Big)
\nonumber\\
&&\hspace{1.2cm}
+\Big({x_{_{\chi_\alpha^0}}\over x_{_{\rm w}}}\Big)^{1/2}
F_4(x_{_{\rm w}},x_{_{\rm w}},x_{_{\chi_\alpha^0}},x_{_{\chi_\beta^\pm}})
\Im\Big({\cal G}^L_{_{\beta\alpha}}\zeta^R_{_{\alpha\beta}}
+{\cal G}^R_{_{\beta\alpha}}\zeta^L_{_{\alpha\beta}}\Big)
\nonumber\\
&&\hspace{1.2cm}
+\Big({x_{_{\chi_\beta^\pm}}\over x_{_{\rm w}}}\Big)^{1/2}
F_5(x_{_{\rm w}},x_{_{\rm w}},x_{_{\chi_\alpha^0}},x_{_{\chi_\beta^\pm}})
\Im\Big({\cal G}^L_{_{\beta\alpha}}\zeta^L_{_{\alpha\beta}}
-{\cal G}^R_{_{\beta\alpha}}\zeta^R_{_{\alpha\beta}}\Big)
\nonumber\\
&&\hspace{1.2cm}
+\Big({x_{_{\chi_\alpha^0}}\over x_{_{\rm w}}}\Big)^{1/2}
F_6(x_{_{\rm w}},x_{_{\rm w}},x_{_{\chi_\alpha^0}},x_{_{\chi_\beta^\pm}})
\Im\Big({\cal G}^L_{_{\beta\alpha}}\zeta^R_{_{\alpha\beta}}
-{\cal G}^R_{_{\beta\alpha}}\zeta^L_{_{\alpha\beta}}\Big)\Bigg\}\;.
\label{MED-W-G}
\end{eqnarray}
Using the asymptotic formulations of form factors $T_{4,5,6,7}$ collected in appendix,
we can simplify the expressions of Eq.\ref{MED-W-G}
in the limit $x_{_{\chi_\alpha^0}},x_{_{\chi_\beta^\pm}}\gg x_{_{\rm w}}$.

The contributions from those above sectors to effective Lagrangian
do not contain ultraviolet divergence. In the pieces discussed below,
the coefficients of high dimensional operators in effective Lagrangian
contain ultraviolet divergence that is caused by the divergent
subdiagrams. In order to obtain physical predictions of lepton
MDMs and EDMs, it is necessary to adopt a concrete renormalization
scheme removing the ultraviolet divergence. In literature, the on-shell renormalization
scheme is adopted frequently to subtract the ultraviolet
divergence which appears in the radiative electroweak corrections
\cite{onshell}. As an over-subtract scheme, the counter terms
include some finite terms which originate from those renormalization conditions
in the on-shell scheme beside the ultraviolet divergence to cancel
the corresponding ultraviolet divergence in the bare Lagrangian. In the concrete
calculation performed here, we apply this scheme to subtract the ultraviolet divergence
caused by the divergent subdiagrams.

\subsection{The effective Lagrangian from the $ZZ$ sector}
\indent\indent
The self energy of $Z$ gauge boson composed of a closed chargino loop
induces the ultraviolet divergence in the Wilson coefficients of effective
Lagrangian. Generally, the unrenormalized self energy of the weak gauge boson
$Z$ can be written as
\begin{eqnarray}
&&\Sigma_{_{\mu\nu}}^{\rm Z}(p)=\Lambda^2A_0^zg_{\mu\nu}+\Big(A_1^z+{p^2\over\Lambda^2}A_2^z\Big)
(p^2g_{\mu\nu}-p_\mu p_\nu)+\Big(B_1^z+{p^2\over\Lambda^2}B_2^z\Big)p_\mu p_\nu\;.
\label{eq-z1}
\end{eqnarray}
Correspondingly, the counter terms are given as
\begin{eqnarray}
&&\Sigma_{_{\mu\nu}}^{\rm ZC}(p)=-(\delta m_{_{\rm z}}^2+m_{_{\rm z}}^2\delta Z_{_{\rm z}})g_{\mu\nu}
-\delta Z_{_{\rm z}}(p^2g_{\mu\nu}-p_\mu p_\nu)\;.
\label{eq-z2}
\end{eqnarray}
The renormalized self energy is given by
\begin{eqnarray}
&&\hat{\Sigma}_{_{\mu\nu}}^{\rm Z}(p)=\Sigma_{_{\mu\nu}}^{\rm Z}(p)
+\Sigma_{_{\mu\nu}}^{\rm ZC}(p)\;.
\label{eq-z3}
\end{eqnarray}
For on-shell external gauge boson $Z$, we have \cite{onshell}
\begin{eqnarray}
&&\hat{\Sigma}_{_{\mu\nu}}^{\rm Z}(p)\epsilon^\nu(p)\Big|_{p^2=m_{_{\rm z}}^2}=0
\;,\nonumber\\
&&\lim\limits_{p^2\rightarrow m_{_{\rm z}}^2}{1\over p^2-m_{_{\rm z}}^2}
\hat{\Sigma}_{_{\mu\nu}}^{\rm Z}(p)\epsilon^\nu(p)=\epsilon_{_\mu}(p)\;,
\label{eq-z4}
\end{eqnarray}
where $\epsilon(p)$ is the polarization vector of $Z$ gauge boson.
From Eq. (\ref{eq-z4}), we get the counter terms
\begin{eqnarray}
&&\delta Z_{_{\rm z}}=A_1^z+{m_{_{\rm z}}^2\over\Lambda^2}A_2^z=A_1^z+x_{_{\rm z}}A_2^z\;,
\nonumber\\
&&\delta m_{_{\rm z}}^2=A_0^z\Lambda^2-m_{_{\rm z}}^2\delta Z_{_{\rm z}}\;.
\label{eq-z5}
\end{eqnarray}

\begin{figure}[t]
\setlength{\unitlength}{1mm}
\begin{center}
\begin{picture}(0,40)(0,0)
\put(-60,-100){\includegraphics{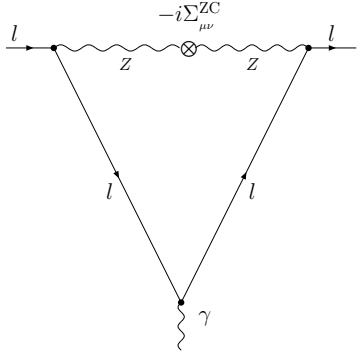}}
\end{picture}
\caption[]{The counter term diagram to cancel the ultraviolet caused
by the self energy of $Z$ boson.}
\label{fig2}
\end{center}
\end{figure}

Accordingly, the effective Lagrangian originating from the counter term
diagram (Fig.\ref{fig2}) can be formulated as
\begin{eqnarray}
&&\delta{\cal L}_{_{ZZ}}^{C}=
-{e^4\over12(4\pi)^2s_{_{\rm w}}^4c_{_{\rm w}}^4\Lambda^2}(4\pi x_{_{\rm R}})^{2\varepsilon}
{\Gamma^2(1+\varepsilon)\over(1-\varepsilon)^2}\Bigg\{\Big(\xi^L_{_{\beta\alpha}}\xi^L_{_{\alpha\beta}}
+\xi^R_{_{\beta\alpha}}\xi^R_{_{\alpha\beta}}\Big)\Big[
-{1\over\varepsilon}{x_{_{\chi_\alpha^\pm}}+x_{_{\chi_\beta^\pm}}\over x_{_{\rm z}}^2}
\nonumber\\
&&\hspace{1.6cm}
+{5(x_{_{\chi_\alpha^\pm}}+x_{_{\chi_\beta^\pm}})\over12x_{_{\rm z}}^2}
+{\varrho_{_{2,1}}(x_{_{\chi_\alpha^\pm}},x_{_{\chi_\beta^\pm}})\over x_{_{\rm z}}^2}
+{5\over12x_{_{\rm z}}}+{x_{_{\chi_\alpha^\pm}}+x_{_{\chi_\beta^\pm}}\over x_{_{\rm z}}^2}
\ln x_{_{\rm R}}\Big]
\nonumber\\
&&\hspace{1.6cm}
+2(x_{_{\chi_\alpha^\pm}}x_{_{\chi_\beta^\pm}})^{1/2}\Big(\xi^L_{_{\beta\alpha}}
\xi^R_{_{\alpha\beta}}+\xi^R_{_{\beta\alpha}}\xi^L_{_{\alpha\beta}}\Big)
\Big[{1\over\varepsilon x_{_{\rm z}}^2}
-{\varrho_{_{1,1}}(x_{_{\chi_\alpha^\pm}},x_{_{\chi_\beta^\pm}})\over x_{_{\rm z}}^2}
+{1\over12x_{_{\rm z}}^2}-{\ln x_{_{\rm R}}\over x_{_{\rm z}}^2}\Big]\Bigg\}
\nonumber\\
&&\hspace{1.6cm}\times
\Big[\Big(T_{_f}^Z-Q_{_f}s_{_{\rm w}}^2\Big)^2({\cal O}_{_2}^-+{\cal O}_{_3}^-)
+Q_{_f}^2s_{_{\rm w}}^4({\cal O}_{_2}^++{\cal O}_{_3}^+)\Big]
\nonumber\\
&&\hspace{1.6cm}
+{e^4\over4(4\pi)^2s_{_{\rm w}}^4c_{_{\rm w}}^4\Lambda^2}(4\pi x_{_{\rm R}})^{2\varepsilon}
{\Gamma^2(1+\varepsilon)\over(1-\varepsilon)^2}\Bigg\{\Big(\xi^L_{_{\beta\alpha}}\xi^L_{_{\alpha\beta}}
+\xi^R_{_{\beta\alpha}}\xi^R_{_{\alpha\beta}}\Big)\Big[
{1\over\varepsilon}{x_{_{\chi_\alpha^\pm}}+x_{_{\chi_\beta^\pm}}\over x_{_{\rm z}}^2}
\nonumber\\
&&\hspace{1.6cm}
-{\varrho_{_{2,1}}(x_{_{\chi_\alpha^\pm}},x_{_{\chi_\beta^\pm}})\over x_{_{\rm z}}^2}
-{x_{_{\chi_\alpha^\pm}}+x_{_{\chi_\beta^\pm}}\over x_{_{\rm z}}^2}({7\over2}+\ln x_{_l}-\ln x_{_{\rm z}})
+{1\over4x_{_{\rm z}}}-{x_{_{\chi_\alpha^\pm}}+x_{_{\chi_\beta^\pm}}\over x_{_{\rm z}}^2}
\ln x_{_{\rm R}}\Big]
\nonumber\\
&&\hspace{1.6cm}
+2(x_{_{\chi_\alpha^\pm}}x_{_{\chi_\beta^\pm}})^{1/2}\Big(\xi^L_{_{\beta\alpha}}
\xi^R_{_{\alpha\beta}}+\xi^R_{_{\beta\alpha}}\xi^L_{_{\alpha\beta}}\Big)
\Big[-{1\over\varepsilon x_{_{\rm z}}^2}
+{\varrho_{_{1,1}}(x_{_{\chi_\alpha^\pm}},x_{_{\chi_\beta^\pm}})\over x_{_{\rm z}}^2}
\nonumber\\
&&\hspace{1.6cm}
+{1\over x_{_{\rm z}}^2}(3+\ln x_{_l}-\ln x_{_{\rm z}})
+{\ln x_{_{\rm R}}\over x_{_{\rm z}}^2}\Big]\Bigg\}
Q_{_f}s_{_{\rm w}}^2\Big(T_{_f}^Z-Q_{_l}s_{_{\rm w}}^2\Big)
({\cal O}_{_6}^-+{\cal O}_{_6}^+)+\cdots\;.
\label{counter-eff}
\end{eqnarray}
Here, $\varepsilon=2-D/2$ with $D$ representing the time-space dimension,
and $x_{_{\rm R}}=\Lambda_{_{\rm RE}}^2/\Lambda^2$ ($\Lambda_{_{\rm RE}}$
denotes the renormalization scale).

As a result of the preparation mentioned above, we can add the contributions from
the counter term diagram to cancel the corresponding ultraviolet
divergence in bare effective Lagrangian. The resulted theoretical predictions
on the lepton MDMs and EDMs are respectively written as
\begin{eqnarray}
&&a_{l,\chi^\pm}^{ZZ}=
-{e^4m_{_l}^2\over(4\pi)^4s_{_{\rm w}}^4c_{_{\rm w}}^4\Lambda^2}\Bigg\{
\Big(|\xi^L_{_{\alpha\beta}}|^2+|\xi^R_{_{\alpha\beta}}|^2\Big)
\Big[\Big(T_{_f}^Z-Q_{_f}s_{_{\rm w}}^2\Big)^2+Q_{_f}^2s_{_{\rm w}}^4\Big]
\nonumber\\
&&\hspace{1.2cm}\times
\Bigg[{Q_{_f}\over3}\Big(T_5(x_{_{\rm z}},x_{_{\chi_\alpha^\pm}},x_{_{\chi_\beta^\pm}})
+{x_{_{\chi_\alpha^\pm}}+x_{_{\chi_\beta^\pm}}\over x_{_{\rm z}}^2}\ln x_{_{\rm R}}\Big)
+{1\over4}T_4(x_{_{\rm z}},x_{_{\chi_\alpha^\pm}},x_{_{\chi_\beta^\pm}})\Bigg]
\nonumber\\
&&\hspace{1.2cm}
+{1\over8}\Big(|\xi^L_{_{\alpha\beta}}|^2-|\xi^R_{_{\alpha\beta}}|^2\Big)
\Big[\Big(T_{_f}^Z-Q_{_f}s_{_{\rm w}}^2\Big)^2-Q_{_f}^2s_{_{\rm w}}^4\Big]
T_{6}(x_{_{\rm z}},x_{_{\chi_\alpha^\pm}},x_{_{\chi_\beta^\pm}})
\nonumber\\
&&\hspace{1.2cm}
-\Re(\xi^L_{_{\alpha\beta}}\xi^R_{_{\beta\alpha}})
\Big[\Big(T_{_f}^Z-Q_{_f}s_{_{\rm w}}^2\Big)^2+Q_{_f}^2s_{_{\rm w}}^4\Big]
(x_{_{\chi_\alpha^\pm}}x_{_{\chi_\beta^\pm}})^{1/2}
\nonumber\\
&&\hspace{1.2cm}\times
\Bigg[{1\over4}T_{7}(x_{_{\rm z}},x_{_{\chi_\alpha^\pm}},x_{_{\chi_\beta^\pm}})
+{4Q_{_f}\over3x_{_{\rm z}}^2}\ln{x_{_{\rm z}}\over x_{_{\rm R}}}-{7Q_{_f}\over3x_{_{\rm z}}^2}\Bigg]
\nonumber\\
&&\hspace{1.2cm}
-\Big(|\xi^L_{_{\alpha\beta}}|^2+|\xi^R_{_{\alpha\beta}}|^2\Big)s_{_{\rm w}}^2
\Big(T_{_f}^Z-Q_{_f}s_{_{\rm w}}^2\Big)
\Bigg[{Q_{_f}\over4}T_{9}(x_{_{\rm z}},x_{_{\chi_\alpha^\pm}},x_{_{\chi_\beta^\pm}})
\nonumber\\
&&\hspace{1.2cm}
-{Q_{_f}^2\over4x_{_{\rm z}}}+{Q_{_f}^2\over x_{_{\rm z}}^2}(2-\ln{x_{_{\rm z}}
\over x_{_{\rm R}}})(x_{_{\chi_\alpha^\pm}}+x_{_{\chi_\beta^\pm}})
-{Q_{_f}^2\over2x_{_{\rm z}}^2}(x_{_{\chi_\alpha^\pm}}\ln x_{_{\chi_\alpha^\pm}}
+x_{_{\chi_\beta^\pm}}\ln x_{_{\chi_\beta^\pm}})
\nonumber\\
&&\hspace{1.2cm}
+{Q_{_f}^2\over2x_{_{\rm z}}^2}\cdot(\varrho_{_{2,1}}(x_{_{\chi_\alpha^\pm}},x_{_{\chi_\beta^\pm}})
-x_{_{\chi_\alpha^\pm}}x_{_{\chi_\beta^\pm}}\varrho_{_{0,1}}(x_{_{\chi_\alpha^\pm}},x_{_{\chi_\beta^\pm}}))\Bigg]
\nonumber\\
&&\hspace{1.2cm}
-4Q_{_f}^2\Re(\xi^L_{_{\alpha\beta}}\xi^R_{_{\beta\alpha}})
s_{_{\rm w}}^2\Big(T_{_f}^Z-Q_{_f}s_{_{\rm w}}^2\Big)
(x_{_{\chi_\alpha^\pm}}x_{_{\chi_\beta^\pm}})^{1/2}
{2-\ln x_{_{\rm z}}+\ln x_{_{\rm R}}\over x_{_{\rm z}}^2}\Bigg\}\;,
\nonumber\\
&&d_{l,\chi^\pm}^{ZZ}=
{e^5m_{_l}\over(4\pi)^4s_{_{\rm w}}^4c_{_{\rm w}}^4\Lambda^2}\cdot
\Im(\xi^L_{_{\alpha\beta}}\xi^R_{_{\beta\alpha}})
(x_{_{\chi_\alpha^\pm}}x_{_{\chi_\beta^\pm}})^{1/2}\Bigg\{
Q_{_f}s_{_{\rm w}}^2\Big(T_{_f}^Z-Q_{_f}s_{_{\rm w}}^2\Big)
\nonumber\\
&&\hspace{1.2cm}\times
\Big({\partial^2\over\partial x_{_{\rm z}}\partial x_{_{\chi_\beta^\pm}}}
-{\partial^2\over\partial x_{_{\rm z}}\partial x_{_{\chi_\alpha^\pm}}}\Big)
\Big({\Phi(x_{_{\rm z}},x_{_{\chi_\alpha^\pm}},x_{_{\chi_\beta^\pm}})
-\varphi_0(x_{_{\chi_\alpha^\pm}},x_{_{\chi_\beta^\pm}})
\over x_{_{\rm z}}}\Big)
\nonumber\\
&&\hspace{1.2cm}
-{1\over16}\Big[\Big(T_{_f}^Z-Q_{_f}s_{_{\rm w}}^2\Big)^2+Q_{_f}^2s_{_{\rm w}}^4\Big]
T_{8}(x_{_{\rm z}},x_{_{\chi_\alpha^\pm}},x_{_{\chi_\beta^\pm}})\Bigg\}\;.
\label{zz-chargino}
\end{eqnarray}

Because a real photon can not be attached to the internal closed neutralino loop,
the corresponding effective Lagrangian only contains the corrections to the
lepton MDMs:
\begin{eqnarray}
&&a_{l,\chi^0}^{ZZ}=
-{e^4Q_{_f}m_{_l}^2\over(4\pi)^4s_{_{\rm w}}^4c_{_{\rm w}}^4\Lambda^2}
\Bigg\{-{1\over3}\Big(|\eta^L_{_{\alpha\beta}}|^2+|\eta^R_{_{\alpha\beta}}|^2\Big)\Big[
\Big(T_{_f}^Z-Q_{_f}s_{_{\rm w}}^2\Big)^2
+Q_{_f}^2s_{_{\rm w}}^4\Big]
\nonumber\\
&&\hspace{1.2cm}\times
\Big(T_5(x_{_{\rm z}},x_{_{\chi_\alpha^0}},x_{_{\chi_\beta^0}})
+{x_{_{\chi_\alpha^0}}+x_{_{\chi_\beta^0}}\over x_{_{\rm z}}^2}\ln x_{_{\rm R}}\Big)
\nonumber\\
&&\hspace{1.2cm}
+{1\over3}\Re(\eta^L_{_{\alpha\beta}}\eta^R_{_{\beta\alpha}})
\Big[\Big(T_{_f}^Z-Q_{_f}s_{_{\rm w}}^2\Big)^2+Q_{_f}^2s_{_{\rm w}}^4\Big]
(x_{_{\chi_\alpha^0}}x_{_{\chi_\beta^0}})^{1/2}
\Bigg[{4\over x_{_{\rm z}}^2}\ln{x_{_{\rm z}}\over x_{_{\rm R}}}-{7\over x_{_{\rm z}}^2}\Bigg]
\nonumber\\
&&\hspace{1.2cm}
+{1\over2 x_{_{\rm z}}^2}\Big(|\eta^L_{_{\alpha\beta}}|^2+|\eta^R_{_{\alpha\beta}}|^2\Big)Q_{_f}s_{_{\rm w}}^2
\Big(T_{_f}^Z-Q_{_f}s_{_{\rm w}}^2\Big)\Bigg[{x_{_{\rm z}}\over2}
+(x_{_{\chi_\alpha^0}}\ln x_{_{\chi_\alpha^0}}+x_{_{\chi_\beta^0}}\ln x_{_{\chi_\beta^0}})
\nonumber\\
&&\hspace{1.2cm}
-2(x_{_{\chi_\alpha^0}}+x_{_{\chi_\beta^0}})(2-\ln{x_{_{\rm z}}\over x_{_{\rm R}}})
-\varrho_{_{2,1}}(x_{_{\chi_\alpha^0}},x_{_{\chi_\beta^0}})
+x_{_{\chi_\alpha^0}}x_{_{\chi_\beta^0}}\varrho_{_{0,1}}
(x_{_{\chi_\alpha^0}},x_{_{\chi_\beta^0}})\Bigg]
\nonumber\\
&&\hspace{1.2cm}
-4Q_{_f}\Re(\eta^L_{_{\alpha\beta}}\eta^R_{_{\beta\alpha}})
s_{_{\rm w}}^2\Big(T_{_f}^Z-Q_{_f}s_{_{\rm w}}^2\Big)
(x_{_{\chi_\alpha^0}}x_{_{\chi_\beta^0}})^{1/2}
{2-\ln x_{_{\rm z}}+\ln x_{_{\rm R}}\over x_{_{\rm z}}^2}\Bigg\}
\label{zz-neutralino}
\end{eqnarray}
with
\begin{eqnarray}
&&\eta^L_{\alpha\beta}={\cal N}_{\alpha4}^\dagger{\cal N}_{4\beta}
\;,\nonumber\\
&&\eta^R_{\alpha\beta}={\cal N}_{\beta3}^\dagger{\cal N}_{3\alpha}
\;,(\alpha,\beta=1,\cdots,4)\;.
\label{coupling-eta}
\end{eqnarray}
We can also simplify Eq.(\ref{zz-chargino}) and Eq.(\ref{zz-neutralino})
using the asymptotic expressions of $T_8\sim T_{13}$ in the limit $
x_{_{\chi_\alpha^\pm}},\;x_{_{\chi_\beta^\pm}},\;x_{_{\chi_\alpha^0}},\;x_{_{\chi_\beta^0}}
\gg x_{_{\rm z}}$.  The concrete expressions of $T_8\sim T_{13}$ can be found
 in appendix.

\subsection{The effective Lagrangian from the $WW$ sector}
\indent\indent
Similarly, the self energy of $W$ gauge boson composed of a closed chargino-neutralino loop
induces the ultraviolet divergence in the Wilson coefficients of effective
Lagrangian. Accordingly, the unrenormalized $W$ self energy is expressed as
\begin{eqnarray}
&&\Sigma_{_{\mu\nu}}^{\rm W}(p)=\Lambda^2A_0^{\rm w}g_{\mu\nu}+\Big(A_1^{\rm w}
+{p^2\over\Lambda^2}A_2^{\rm w}\Big)(p^2g_{\mu\nu}-p_\mu p_\nu)
+\Big(B_1^{\rm w}+{p^2\over\Lambda^2}B_2^{\rm w}\Big)p_\mu p_\nu\;.
\label{eq-w1}
\end{eqnarray}
The corresponding counter terms are given as
\begin{eqnarray}
&&\Sigma_{_{\mu\nu}}^{\rm WC}(p)=-(\delta m_{_{\rm w}}^2+m_{_{\rm w}}^2\delta Z_{_{\rm w}})g_{\mu\nu}
-\delta Z_{_{\rm w}}(p^2g_{\mu\nu}-p_\mu p_\nu)\;.
\label{eq-w2}
\end{eqnarray}

\begin{figure}[t]
\setlength{\unitlength}{1mm}
\begin{center}
\begin{picture}(0,100)(0,0)
\put(-60,-40){\includegraphics{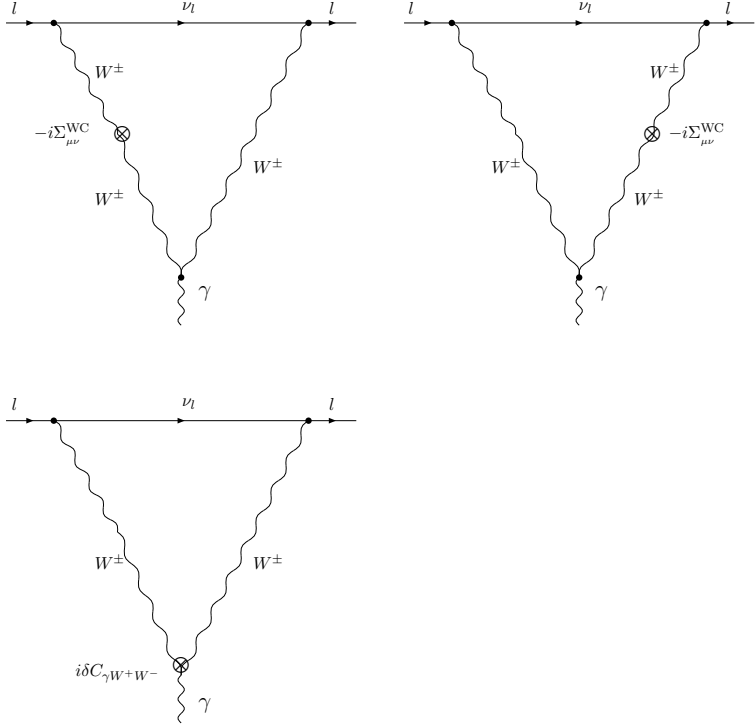}}
\end{picture}
\caption[]{The counter term diagram to cancel the ultraviolet caused
by the self energy of $W$ boson and electroweak radiative corrections to
$\gamma W^+W^-$ vertex.}
\label{fig3}
\end{center}
\end{figure}

The renormalized self energy is given by
\begin{eqnarray}
&&\hat{\Sigma}_{_{\mu\nu}}^{\rm W}(p)=\Sigma_{_{\mu\nu}}^{\rm W}(p)+\Sigma_{_{\mu\nu}}^{\rm WC}(p)
\label{eq-w3}
\end{eqnarray}
For on-shell external gauge boson $W^\pm$, we have \cite{onshell}
\begin{eqnarray}
&&\hat{\Sigma}_{_{\mu\nu}}^{\rm W}(p)\epsilon^\nu(p)\Big|_{p^2=m_{_{\rm w}}^2}=0
\;,\nonumber\\
&&\lim\limits_{p^2\rightarrow m_{_{\rm w}}^2}{1\over p^2-m_{_{\rm w}}^2}
\hat{\Sigma}_{_{\mu\nu}}^{\rm W}(p)\epsilon^\nu(p)=\epsilon_{_\mu}(p)\;,
\label{eq-w4}
\end{eqnarray}
where $\epsilon(p)$ is the polarization vector of $W$ gauge boson.
Inserting Eq. (\ref{eq-w1}) and Eq. (\ref{eq-w2}) into Eq. (\ref{eq-w4}),
we derive the counter terms for the $W$ self energy as
\begin{eqnarray}
&&\delta Z_{_{\rm w}}=A_1^{\rm w}+{m_{_{\rm w}}^2\over\Lambda^2}A_2^{\rm w}
=A_1^{\rm w}+x_{_{\rm z}}A_2^{\rm w}\;,
\nonumber\\
&&\delta m_{_{\rm w}}^2=A_0^{\rm w}\Lambda^2-m_{_{\rm w}}^2\delta Z_{_{\rm w}}\;.
\label{eq-w5}
\end{eqnarray}
Differing from the analysis in the $ZZ$ sector, we should derive the
counter term for the vertex $\gamma W^+W^-$ here since the corresponding coupling
is not zero at tree level. In the nonlinear $R_\xi$ gauge with $\xi=1$,
the counter term for the vertex $\gamma W^+W^-$ is
\begin{eqnarray}
&&i\delta C_{\gamma W^+W^-}=ie\cdot\delta Z_{_{\rm w}}\Big[g_{\mu\nu}(k_1-k_2)_\rho
+g_{\nu\rho}(k_2-k_3)_\mu+g_{\rho\mu}(k_3-k_1)_\nu\Big]\;,
\label{eq-w6}
\end{eqnarray}
where $k_i\;(i=1,\;2,\;3)$ denote the injection momenta of $W^\pm$ and photon,
and $\mu,\;\nu,\;\rho$ denote the corresponding Lorentz indices respectively.

We present the counter term diagrams to cancel the ultraviolet divergence contained in
the bare effective Lagrangian from $WW$
sector in Fig.\ref{fig3}, and we can verify that the sum of corresponding amplitude
satisfies the Ward identity required by the QED gauge invariance obviously.
Accordingly, the effective Lagrangian originating from the counter
term diagrams can be written as

\begin{eqnarray}
&&\delta{\cal L}_{_{WW}}^C=
{e^4\over(4\pi)^2s_{_{\rm w}}^4\Lambda^2Q_{_f}}(4\pi x_{_{\rm R}})^{2\varepsilon}
{\Gamma^2(1+\varepsilon)\over(1-\varepsilon)^2}\Bigg\{\Big(\zeta^{L*}_{_{\alpha\beta}}
\zeta^L_{_{\alpha\beta}}+\zeta^{R*}_{_{\alpha\beta}}\zeta^R_{_{\alpha\beta}}\Big)
\nonumber\\
&&\hspace{1.4cm}\times
\Big[{5\over24x_{_{\rm w}}^2}\Big(-{x_{_{\chi_\alpha^0}}+x_{_{\chi_\beta^\pm}}\over\varepsilon}
-{x_{_{\chi_\alpha^0}}+x_{_{\chi_\beta^\pm}}\over3}+\varrho_{_{2,1}}(x_{_{\chi_\alpha^0}},x_{_{\chi_\beta^\pm}})
\nonumber\\
&&\hspace{1.4cm}
+(x_{_{\chi_\alpha^0}}+x_{_{\chi_\beta^\pm}})\ln x_{_{\rm R}}\Big)+{11\over36x_{_{\rm w}}}
\Big]({\cal O}_{_2}^-+{\cal O}_{_3}^-)
\nonumber\\
&&\hspace{1.4cm}
+\Big(\zeta^{L*}_{_{\alpha\beta}}\zeta^R_{_{\alpha\beta}}+\zeta^{R*}_{_{\alpha\beta}}\zeta^L_{_{\alpha\beta}}\Big)
(x_{_{\chi_\alpha^0}}x_{_{\chi_\beta^\pm}})^{1/2}\Big[{5\over12x_{_{\rm w}}^2}\Big(
{1\over\varepsilon}+{5\over6}-\varrho_{_{1,1}}(x_{_{\chi_\alpha^0}},x_{_{\chi_\beta^\pm}})
\nonumber\\
&&\hspace{1.4cm}
-\ln x_{_{\rm R}}\Big)\Big]
({\cal O}_{_2}^-+{\cal O}_{_3}^-)\Bigg\}+\cdots\;.
\label{w-counter}
\end{eqnarray}

Finally, we get the renormalized effective Lagrangian from the $WW$ sector:
\begin{eqnarray}
&&{\cal L}_{_{WW}}=
-{e^4\over48(4\pi)^2s_{_{\rm w}}^4Q_{_f}\Lambda^2}
\Big(\zeta^{L*}_{_{\alpha\beta}}\zeta^L_{_{\alpha\beta}}
+\zeta^{R*}_{_{\alpha\beta}}\zeta^R_{_{\alpha\beta}}\Big)
\Big[T_{10}(x_{_{\rm w}},x_{_{\chi_\alpha^0}},x_{_{\chi_\beta^\pm}})
\nonumber\\
&&\hspace{1.4cm}
+{10\over x_{_{\rm w}}^2}(x_{_{\chi_\alpha^0}}+x_{_{\chi_\beta^\pm}})\ln x_{_{\rm R}}
\Big]({\cal O}_{_2}^-+{\cal O}_{_3}^-)
\nonumber\\
&&\hspace{1.4cm}
-{e^4\over16(4\pi)^2s_{_{\rm w}}^4Q_{_f}\Lambda^2}
\Big(\zeta^{L*}_{_{\alpha\beta}}\zeta^L_{_{\alpha\beta}}
-\zeta^{R*}_{_{\alpha\beta}}\zeta^R_{_{\alpha\beta}}\Big)
T_{11}(x_{_{\rm w}},x_{_{\chi_\alpha^0}},x_{_{\chi_\beta^\pm}})
({\cal O}_{_2}^-+{\cal O}_{_3}^-)
\nonumber\\
&&\hspace{1.4cm}
-{e^4(x_{_{\chi_\alpha^0}}x_{_{\chi_\beta^\pm}})^{1/2}\over48(4\pi)^2s_{_{\rm w}}^4Q_{_f}\Lambda^2}
\Big(\zeta^{L*}_{_{\alpha\beta}}
\zeta^R_{_{\alpha\beta}}+\zeta^{R*}_{_{\alpha\beta}}\zeta^L_{_{\alpha\beta}}\Big)
\Big[T_{12}(x_{_{\rm w}},x_{_{\chi_\alpha^0}},x_{_{\chi_\beta^\pm}})
-{20\over x_{_{\rm w}}^2}\ln x_{_{\rm R}}\Big]({\cal O}_{_2}^-+{\cal O}_{_3}^-)
\nonumber\\
&&\hspace{1.4cm}
-{e^4(x_{_{\chi_\alpha^0}}x_{_{\chi_\beta^\pm}})^{1/2}\over16(4\pi)^2s_{_{\rm w}}^4Q_{_f}\Lambda^2}
\Big(\zeta^{R*}_{_{\alpha\beta}}
\zeta^L_{_{\alpha\beta}}-\zeta^{L*}_{_{\alpha\beta}}\zeta^R_{_{\alpha\beta}}\Big)
T_{13}(x_{_{\rm w}},x_{_{\chi_\alpha^0}},x_{_{\chi_\beta^\pm}})
({\cal O}_{_2}^--{\cal O}_{_3}^-)\;.
\label{ww}
\end{eqnarray}
Correspondingly, the resulted lepton MDMs and EDMs are respectively formulated
as
\begin{eqnarray}
&&a_{l}^{WW}=
-{e^4m_{_l}^2\over12(4\pi)^4s_{_{\rm w}}^4\Lambda^2}
\Big(|\zeta^L_{_{\alpha\beta}}|^2+|\zeta^R_{_{\alpha\beta}}|^2\Big)
\Big[T_{10}(x_{_{\rm w}},x_{_{\chi_\alpha^0}},x_{_{\chi_\beta^\pm}})
\nonumber\\
&&\hspace{1.4cm}
+{10\over x_{_{\rm w}}^2}(x_{_{\chi_\alpha^0}}+x_{_{\chi_\beta^\pm}})\ln x_{_{\rm R}}
-{32\over x_{_{\rm w}}}\ln x_{_{\rm R}}\Big]
\nonumber\\
&&\hspace{1.4cm}
-{e^4m_{_l}^2\over4(4\pi)^4s_{_{\rm w}}^4\Lambda^2}
\Big(|\zeta^L_{_{\alpha\beta}}|^2-|\zeta^R_{_{\alpha\beta}}|^2\Big)
T_{11}(x_{_{\rm w}},x_{_{\chi_\alpha^0}},x_{_{\chi_\beta^\pm}})
\nonumber\\
&&\hspace{1.4cm}
-{e^4m_{_l}^2(x_{_{\chi_\alpha^0}}x_{_{\chi_\beta^\pm}})^{1/2}\over6(4\pi)^4s_{_{\rm w}}^4\Lambda^2}
\Re(\zeta^{R*}_{_{\alpha\beta}}\zeta^L_{_{\alpha\beta}})
\Big[T_{12}(x_{_{\rm w}},x_{_{\chi_\alpha^0}},x_{_{\chi_\beta^\pm}})
-{20\over x_{_{\rm w}}^2}\ln x_{_{\rm R}}\Big]\;,
\nonumber\\
&&d_{l}^{WW}=
-{e^5m_{_l}(x_{_{\chi_\alpha^0}}x_{_{\chi_\beta^\pm}})^{1/2}\over4(4\pi)^4s_{_{\rm w}}^4\Lambda^2}
\Im(\zeta^{R*}_{_{\alpha\beta}}\zeta^L_{_{\alpha\beta}})
T_{13}(x_{_{\rm w}},x_{_{\chi_\alpha^0}},x_{_{\chi_\beta^\pm}})\;.
\label{MEDM-ww}
\end{eqnarray}

\section{Numerical results and discussion\label{sec3}}
\indent\indent
With the theoretical formulae derived in
previous section, we numerically analyze the dependence of the muon MDM
and the electron EDM on the supersymmetric parameters in the split scenario
here. In particular, we will present the dependence of the muon MDM
and the electron EDM on the supersymmetric $CP$ phases in some detail. Within
three standard error deviations, the present experimental data can
tolerate new physics corrections to the muon MDM as
$-10\times10^{-10}<\Delta a_\mu <52\times10^{-10}$. Since the
neutralinos $\chi_{\alpha}^0\;(\alpha=1,\;2,\;3,\;4)$
and charginos $\chi_{\alpha}^\pm\;(\alpha=1,\;2)$ appear
as the internal intermediate particles in the two-loop diagrams
which are investigated in this work, the corrections of these
diagrams will be suppressed strongly when the masses of neutralinos and
charginos are much higher than the electroweak scale\cite{heinemeyer2}.
To investigate if those diagrams can result in concrete corrections to
the muon MDM and electron EDM, we choose a suitable supersymmetric parameter region where
the masses of neutralinos and charginos are lying in the
range $M_{_{\chi}}<500\;{\rm GeV}$. Without losing too much generality,
we assume the supersymmetric parameters satisfying $|m_1|=|m_2|$ in this work.
In split SUSY, only the CP violating phases $\arg(\mu_{_H}),\;\arg(m_1)$,
and $\arg(m_2)$ have substantive contributions to the lepton MDMs and EDMs.
As for other CP violating phases, either they do not contribute to the lepton MDMs and EDMs
or their corrections can be neglected safely. Moreover,
the existence of a CP-even SM like Higgs with mass around $100-250\;{\rm GeV}$
sets a strong constraint on the parameter space of the employed model.
To address this problem, they argue that a fine tuning in the Higgs potential is required \cite{Dress}.
Admitting this fine tuning among high energy scale parameters, one no longer
worries about the constraint from Higgs sector. On the other hand, the CP violation would
cause changes to the neutral-Higgs-quark coupling, the neutral Higgs-gauge-boson
coupling and the self-coupling of Higgs boson. The present
experimental lower bound on the mass of the lightest Higgs bosons
is relaxed to 60 GeV \cite{Pilaftsis}. The scope of Higgs mass is chosen around
$60-250\;{\rm GeV}$ in our numerical analysis because of the above reasons.
In fact, we find that the lepton MDMs and EDMs weakly
depend on the mass of lightest Higgs by scanning
the parameter space. In the following discussion, we choose
the mass of light Higgs as $m_h=120\;{\rm GeV}$.

\begin{figure}[t]
\setlength{\unitlength}{1mm}
\begin{center}
\begin{picture}(0,80)(0,0)
\put(-50,-15){\includegraphics{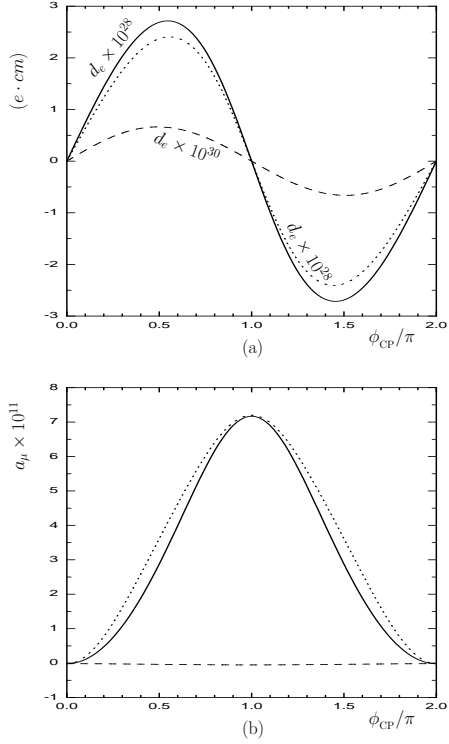}}
\end{picture}
\caption[]{The supersymmetric corrections to the electron EDM $d_e$ and muon MDM $a_\mu$
vary with the CP violating phase $\phi_{_{\rm CP}}$
when $|\mu_{_H}|=|m_1|=|m_2|=200\;{\rm GeV}$ and $\tan\beta=5$. Where
the dash lines stand for the corrections with $\phi_{_{\rm CP}}=\arg(m_1)$
and $\arg(m_2)=\arg(\mu_{_H})=0$,
the dot lines stand for the corrections with $\phi_{_{\rm CP}}=\arg(m_2)$
and $\arg(m_1)=\arg(\mu_{_H})=0$,
and the solid lines stand for the corrections with $\phi_{_{\rm CP}}=\arg(\mu_{_H})$
and $\arg(m_1)=\arg(m_2)=0$,
respectively.}
\label{fig4}
\end{center}
\end{figure}

Taking $|\mu_{_H}|=|m_1|=|m_2|=200\;{\rm GeV}$ and $\tan\beta=5$,
we plot the electron EDM $d_e$ and muon MDM $a_\mu$ versus the CP
phases $\phi_{_{\rm CP}}=\arg(m_1),\;\arg(m_2),\;\arg(\mu_{_H})$
separately in Fig.\ref{fig4}. When $\arg(m_2)=\arg(\mu_{_H})=0$,
there is cancellation among the dominant contributions to $a_\mu$
that originate from the "$\gamma Z$" and "$ZZ$" sectors respectively.
As $\phi_{_{\rm CP}}=\arg(\mu_{_H})$ and $\arg(m_1)=\arg(m_2)=0$,
the absolute values of supersymmetric corrections to
the electron EDM $d_e$ (solid line in Fig.\ref{fig4}(a)) exceed $2.5\times10^{-28}\;e\cdot cm$
at the largest CP violation $\arg(\mu_{_H})=\pi/2,\;3\pi/2$,
which is well below the present experimental upper limit $1.7\times10^{-27}\;e\cdot cm$
\cite{Regan}, but large enough to be detected in next generation experiments.
Correspondingly, the muon MDM $a_\mu$ depends on the CP violating phase
$\phi_{_{\rm CP}}=\arg(\mu_{_H})$ (solid line in Fig.\ref{fig4}(b)) strongly,
the supersymmetric corrections to $a_\mu$ exceed $3\times10^{-11}$ at the
largest CP violation ($\arg(\mu_{_H})=\pi/2,\;3\pi/2$) since
the cancellation mentioned above dissolves now. For the same reason,
the supersymmetric correction to $a_\mu$ surpasses $7\times10^{-11}$
at the CP conservation of $\arg(\mu_{_H})=\pi$.
As $\phi_{_{\rm CP}}=\arg(m_2)$ and $\arg(m_1)=\arg(\mu_{_H})=0$,
the absolute values of supersymmetric corrections to
the electron EDM $d_e$ (solid line in Fig.\ref{fig4}(a)) exceed $2\times10^{-28}\;e\cdot cm$
at the largest CP violation $\arg(\mu_{_H})=\pi/2,\;3\pi/2$,
which is expected to be observed in next
generation experiments with the sensitivity $10^{-29}\;e\cdot cm$ \cite{Kawell}.
Correspondingly, the muon MDM $a_\mu$ depends on the CP violating phase
$\phi_{_{\rm CP}}=\arg(m_2)$ (dot line in Fig.\ref{fig4}(b)) sensitively,
the supersymmetric corrections to $a_\mu$ are about $3\times10^{-11}$
at the largest CP violation ($\arg(\mu_{_H})=\pi/2,\;3\pi/2$) because
the cancellation existing as $\arg(m_2)=\arg(\mu_{_H})=0$ vanishes here.
When $\phi_{_{\rm CP}}=\arg(m_1)$ and $\arg(m_2)=\arg(\mu_{_H})=0$,
the supersymmetric correction to the electron EDM $d_e$ (dash line in
Fig.\ref{fig4}(a)) is below $1\times10^{-30}\;e\cdot cm$,
and very difficult to be detected in near future.
Corresponding to small theoretical prediction on the electron EDM $d_e$,
the muon MDM $a_\mu$ varies with the CP phase $\phi_{_{\rm CP}}=\arg(m_1)$
(dash line Fig.\ref{fig4}(b)) very mildly.

\begin{figure}[t]
\setlength{\unitlength}{1mm}
\begin{center}
\begin{picture}(0,80)(0,0)
\put(-50,-15){\includegraphics{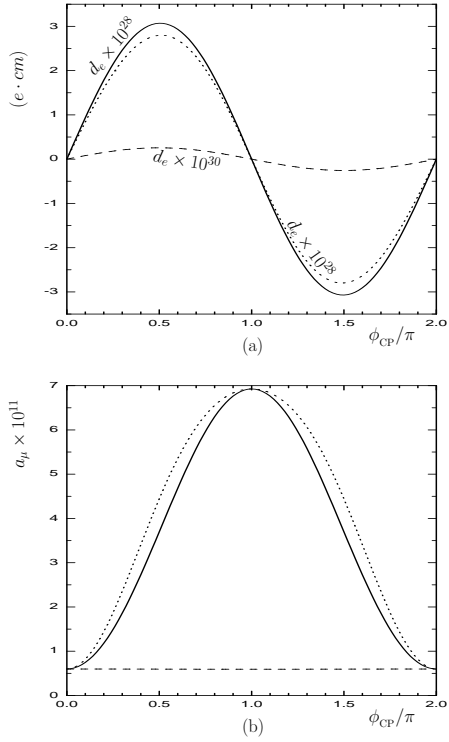}}
\end{picture}
\caption[]{The supersymmetric corrections to the electron EDM $d_e$ and muon MDM $a_\mu$
vary with the CP violating phase $\phi_{_{\rm CP}}$
when $|\mu_{_H}|=|m_1|=|m_2|=200\;{\rm GeV}$ and $\tan\beta=50$. Where
the dash lines stand for the corrections with $\phi_{_{\rm CP}}=\arg(m_1)$
and $\arg(m_2)=\arg(\mu_{_H})=0$,
the dot lines stand for the corrections with $\phi_{_{\rm CP}}=\arg(m_2)$
and $\arg(m_1)=\arg(\mu_{_H})=0$,
and the solid lines stand for the corrections with $\phi_{_{\rm CP}}=\arg(\mu_{_H})$
and $\arg(m_1)=\arg(m_2)=0$,
separately.}
\label{fig5}
\end{center}
\end{figure}

Taking $|\mu_{_H}|=|m_1|=|m_2|=200\;{\rm GeV}$ and $\tan\beta=50$,
we plot the electron EDM $d_e$ and muon MDM $a_\mu$ versus the CP
phases $\phi_{_{\rm CP}}=\arg(m_1),\;\arg(m_2),\;\arg(\mu_{_H})$
separately in Fig.\ref{fig5}.
As $\phi_{_{\rm CP}}=\arg(\mu_{_H})$ and $\arg(m_1)=\arg(m_2)=0$,
the absolute values of supersymmetric corrections to
the electron EDM $d_e$ (solid line in Fig.\ref{fig5}(a)) reach $3\times10^{-28}\;e\cdot cm$
at the largest CP violation $\arg(\mu_{_H})=\pi/2,\;3\pi/2$,
which exceeds the precision of next generation experiments\cite{Kawell}.
Correspondingly, the muon MDM $a_\mu$ depends on the CP violating phase
$\phi_{_{\rm CP}}=\arg(\mu_{_H})$ (solid line in Fig.\ref{fig5}(b)) strongly,
the supersymmetric corrections to $a_\mu$ are about $3\times10^{-11}$
at the largest CP violation ($\arg(\mu_{_H})=\pi/2,\;3\pi/2$) since
the cancellation appearing at $\arg(m_2)=\arg(\mu_{_H})=0$ dissolves here.
As $\phi_{_{\rm CP}}=\arg(m_2)$ and $\arg(m_1)=\arg(\mu_{_H})=0$,
the absolute values of supersymmetric corrections to
the electron EDM $d_e$ (solid line in Fig.\ref{fig5}(a)) exceed $2.5\times10^{-28}\;e\cdot cm$
at the largest CP violation $\arg(\mu_{_H})=\pi/2,\;3\pi/2$,
which is expected to be observed in next
generation experiments with the sensitivity $10^{-29}\;e\cdot cm$.
Correspondingly, the muon MDM $a_\mu$ depends on the CP violating phase
$\phi_{_{\rm CP}}=\arg(m_2)$ (dot line in Fig.\ref{fig5}(b)) steeply,
the supersymmetric corrections to $a_\mu$ are about $4\times10^{-11}$
at the largest CP violation ($\arg(\mu_{_H})=\pi/2,\;3\pi/2$) because the
cancellation existing at $\arg(m_2)=\arg(\mu_{_H})=0$ disappears presently.
When $\phi_{_{\rm CP}}=\arg(m_1)$ and $\arg(m_2)=\arg(\mu_{_H})=0$, the supersymmetric correction to
the electron EDM $d_e$ (dash line in Fig.\ref{fig5}(a)) is less than $1\times10^{-30}\;e\cdot cm$,
and very difficult to be detected in near future.
Corresponding to small theoretical prediction on the electron EDM $d_e$,
the muon MDM $a_\mu$ varies with the CP phase $\phi_{_{\rm CP}}=\arg(m_1)$
(dash line Fig.\ref{fig5}(b)) very slowly.

\begin{figure}[t]
\setlength{\unitlength}{1mm}
\begin{center}
\begin{picture}(0,80)(0,0)
\put(-50,-15){\includegraphics{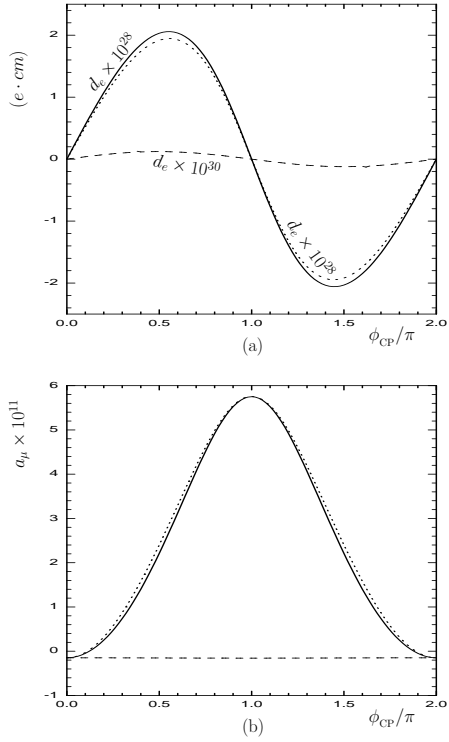}}
\end{picture}
\caption[]{The supersymmetric corrections to the electron EDM $d_e$ and muon MDM $a_\mu$
vary with the CP violating phase $\phi_{_{\rm CP}}$
when $|\mu_{_H}|=|m_1|=|m_2|=300\;{\rm GeV}$ and $\tan\beta=5$. Where
the dash lines stand for the corrections with $\phi_{_{\rm CP}}=\arg(m_1)$
and $\arg(m_2)=\arg(\mu_{_H})=0$,
the dot lines stand for the corrections with $\phi_{_{\rm CP}}=\arg(m_2)$
and $\arg(m_1)=\arg(\mu_{_H})=0$,
and the solid lines stand for the corrections with $\phi_{_{\rm CP}}=\arg(\mu_{_H})$
and $\arg(m_1)=\arg(m_2)=0$, separately.}
\label{fig6}
\end{center}
\end{figure}

Note that the theoretical predictions on the electron EDM $d_e$ and muon MDM $a_\mu$
are not enhanced by large $\tan\beta$ here, this point
can be understood as follows. The $\tan\beta$ enhanced couplings are only contained
in the interactions among the heavy Higgs and down-type fermions (sfermions).
Those heavy Higgs fields include the neutral CP-odd Higgs, the neutral heavy CP-even Higgs,
as well as the charged Higgs. However, those particles are
decoupled from the low energy theory because they are super-heavy under the split assumption.
In other words, our theoretical predictions are not enhanced by large $\tan\beta$ since we
ignore the corrections from those heavy Higgs fields.

Taking $|\mu_{_H}|=|m_1|=|m_2|=300\;{\rm GeV}$ and $\tan\beta=5$,
we plot the electron EDM $d_e$ and muon MDM $a_\mu$ versus the CP
phases $\phi_{_{\rm CP}}=\arg(m_1),\;\arg(m_2),\;\arg(\mu_{_H})$
separately in Fig.\ref{fig6}. As $\arg(m_2)=\arg(\mu_{_H})=0$,
a cancellation exists among the dominant supersymmetric contributions to $a_\mu$
that originate from the "$\gamma Z$" and "$ZZ$" sectors respectively.
When $\phi_{_{\rm CP}}=\arg(\mu_{_H})$ and $\arg(m_1)=\arg(m_2)=0$,
the absolute values of supersymmetric corrections to
the electron EDM $d_e$ (solid line in Fig.\ref{fig6}(a)) approach $2\times10^{-28}\;e\cdot cm$
at the largest CP violation $\arg(\mu_{_H})=\pi/2,\;3\pi/2$,
which is well below the present experimental upper limit $1.7\times10^{-27}\;e\cdot cm$,
but large enough to be detected in next generation experiments with the
precision of $10\times10^{-29}\;e\cdot cm$.
Correspondingly, the muon MDM $a_\mu$ depends on the CP violating phase
$\phi_{_{\rm CP}}=\arg(\mu_{_H})$ (solid line in Fig.\ref{fig6}(b)) strongly,
the supersymmetric corrections to $a_\mu$ exceed $2\times10^{-11}$
at the largest CP violation ($\arg(\mu_{_H})=\pi/2,\;3\pi/2$) since
the cancellation mentioned above dissolves here. For the same reason,
the supersymmetric correction to $a_\mu$ surpasses $5.5\times10^{-11}$
at the CP conservation of $\arg(\mu_{_H})=\pi$.
As $\phi_{_{\rm CP}}=\arg(m_2)$ and $\arg(m_1)=\arg(\mu_{_H})=0$,
the absolute values of supersymmetric corrections to
the electron EDM $d_e$ (solid line in Fig.\ref{fig6}(a)) exceed $2\times10^{-28}\;e\cdot cm$
at the largest CP violation $\arg(\mu_{_H})=\pi/2,\;3\pi/2$,
which is expected to be observed in next generation experiments.
Correspondingly, the muon MDM $a_\mu$ depends on the CP violating phase
$\phi_{_{\rm CP}}=\arg(m_2)$ (dot line in Fig.\ref{fig6}(b)) steeply,
the supersymmetric corrections to $a_\mu$ are about $2\times10^{-11}$
at the largest CP violation ($\arg(\mu_{_H})=\pi/2,\;3\pi/2$) because
the cancellation appearing at $\arg(m_2)=\arg(\mu_{_H})=0$ vanishes now.
When $\phi_{_{\rm CP}}=\arg(m_1)$ and $\arg(m_2)=\arg(\mu_{_H})=0$, the supersymmetric correction to
the electron EDM $d_e$ (dash line in Fig.\ref{fig6}(a)) is below $1\times10^{-30}\;e\cdot cm$,
and very difficult to be detected in near future.
Corresponding to small theoretical prediction on the electron EDM $d_e$,
the muon MDM $a_\mu$ varies with the CP phase $\phi_{_{\rm CP}}=\arg(m_1)$
(dash line Fig.\ref{fig6}(b)) very slowly.

\begin{figure}[t]
\setlength{\unitlength}{1mm}
\begin{center}
\begin{picture}(0,80)(0,0)
\put(-50,-15){\includegraphics{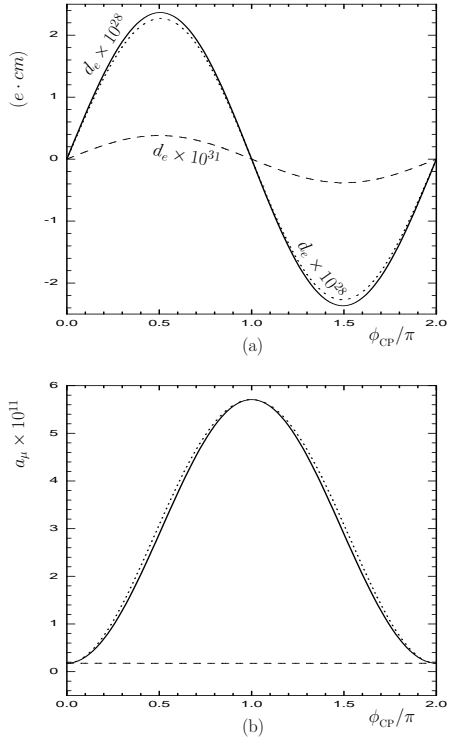}}
\end{picture}
\caption[]{The supersymmetric corrections to the electron EDM $d_e$ and muon MDM $a_\mu$
vary with the CP violating phase $\phi_{_{\rm CP}}$
when $|\mu_{_H}|=|m_1|=|m_2|=300\;{\rm GeV}$ and $\tan\beta=50$. Where
the dash lines stand for the corrections with $\phi_{_{\rm CP}}=\arg(m_1)$
and $\arg(m_2)=\arg(\mu_{_H})=0$,
the dot lines stand for the corrections with $\phi_{_{\rm CP}}=\arg(m_2)$
and $\arg(m_1)=\arg(\mu_{_H})=0$,
and the solid lines stand for the corrections with $\phi_{_{\rm CP}}=\arg(\mu_{_H})$
and $\arg(m_1)=\arg(m_2)=0$, respectively.}
\label{fig7}
\end{center}
\end{figure}

Taking $|\mu_{_H}|=|m_1|=|m_2|=300\;{\rm GeV}$ and $\tan\beta=50$,
we plot the electron EDM $d_e$ and muon MDM $a_\mu$ versus the CP
phases $\phi_{_{\rm CP}}=\arg(m_1),\;\arg(m_2),\;\arg(\mu_{_H})$
separately in Fig.\ref{fig7}.
As $\phi_{_{\rm CP}}=\arg(\mu_{_H})$ and $\arg(m_1)=\arg(m_2)=0$,
the absolute values of supersymmetric corrections to
the electron EDM $d_e$ (solid line in Fig.\ref{fig5}(a)) reach $2.3\times10^{-28}\;e\cdot cm$
at the largest CP violation $\arg(\mu_{_H})=\pi/2,\;3\pi/2$,
which exceeds the precision of next generation experiments.
Correspondingly, the muon MDM $a_\mu$ depends on the CP violating phase
$\phi_{_{\rm CP}}=\arg(\mu_{_H})$ (solid line in Fig.\ref{fig7}(b)) strongly,
the supersymmetric corrections to $a_\mu$ are about $2\times10^{-11}$
at the largest CP violation ($\arg(\mu_{_H})=\pi/2,\;3\pi/2$) since
the cancellation existing at $\arg(m_2)=\arg(\mu_{_H})=0$ dissolves here.
As $\phi_{_{\rm CP}}=\arg(m_2)$ and $\arg(m_1)=\arg(\mu_{_H})=0$,
the absolute values of supersymmetric corrections to
the electron EDM $d_e$ (solid line in Fig.\ref{fig7}(a)) exceed $2.2\times10^{-28}\;e\cdot cm$
at the largest CP violation $\arg(\mu_{_H})=\pi/2,\;3\pi/2$,
which is expected to be observed in next generation experiments.
Correspondingly, the muon MDM $a_\mu$ depends on the CP violating phase
$\phi_{_{\rm CP}}=\arg(m_2)$ (dot line in Fig.\ref{fig7}(b)) steeply,
the supersymmetric corrections to $a_\mu$ are about $2\times10^{-11}$
at the largest CP violation ($\arg(\mu_{_H})=\pi/2,\;3\pi/2$) because
the cancellation mentioned above vanishes also now.
When $\phi_{_{\rm CP}}=\arg(m_1)$ and $\arg(m_2)=\arg(\mu_{_H})=0$, the supersymmetric correction to
the electron EDM $d_e$ (dash line in Fig.\ref{fig7}(a)) is less than
$1\times10^{-31}\;e\cdot cm$, and very difficult to be detected in near future.
Corresponding to small theoretical prediction on the electron EDM $d_e$,
the muon MDM $a_\mu$ varies with the CP phase $\phi_{_{\rm CP}}=\arg(m_1)$
(dash line Fig.\ref{fig7}(b)) very mildly.

\begin{figure}[t]
\setlength{\unitlength}{1mm}
\begin{center}
\begin{picture}(0,80)(0,0)
\put(-50,-15){\includegraphics{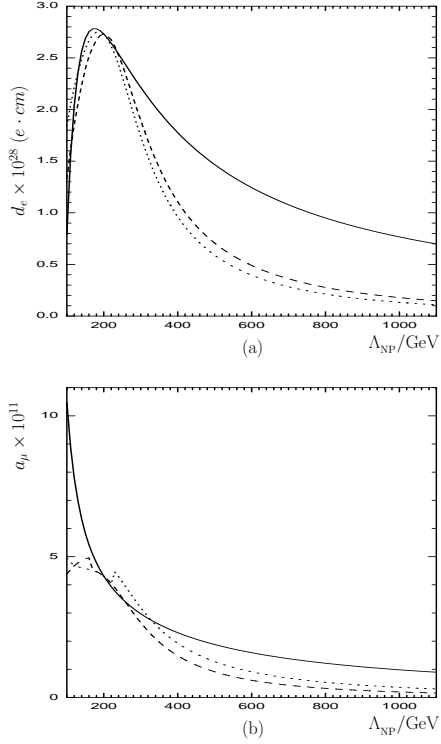}}
\end{picture}
\caption[]{The supersymmetric corrections to the electron EDM $d_e$ and muon MDM $a_\mu$
vary with the energy scale $\Lambda$ when $\tan\beta=20$, and $\arg(\mu_{_H})=\arg(m_1)=0,
\;\arg(m_2)=\pi/2$. Where the dot lines stand for the corrections with
$|\mu_{_H}|=200\;{\rm GeV},|m_1|=|m_2|=\Lambda_{_{\rm NP}}$,
the dash lines stand for the corrections with $|m_1|=|m_2|=200\;{\rm GeV},\;
|\mu_{_H}|=\Lambda_{_{\rm NP}}$, and the solid lines stand for the corrections
with $|\mu_{_H}|=|m_1|=|m_2|=\Lambda_{_{\rm NP}}$, respectively.}
\label{fig8}
\end{center}
\end{figure}

In the numerical analysis presented above, the assumption
$|\mu_{_H}|=|m_1|=|m_2|$ is taken for simplicity. This assumption
on parameter space induces very specific mixing patterns in chargino
and neutralino sectors respectively. In order to investigate the
supersymmetric corrections to $a_\mu$ and $d_e$ without the assumption,
we plot $a_\mu$ and $d_e$ varying with the energy scale of new physics
$\Lambda_{_{\rm NP}}$ when $\arg(\mu_{_H})=\arg(m_1)=0,\;\arg(m_2)=\pi/2$
in Fig.\ref{fig8}. Because the supersymmetric corrections to $a_\mu$
and $d_e$ depend on $\tan\beta$ mildly, we choose a middle value of $\tan\beta=20$.
In Fig.\ref{fig8}(a) and Fig.\ref{fig8}(b), the solid lines stand for
the supersymmetric corrections with $|\mu_{_H}|=|m_1|=|m_2|=\Lambda_{_{\rm NP}}$,
the dot lines stand for the supersymmetric corrections with $|\mu_{_H}|=200\;{\rm GeV},
|m_1|=|m_2|=\Lambda_{_{\rm NP}}$, and
the dash lines stand for the supersymmetric corrections with $|m_1|=|m_2|=200\;{\rm GeV},\;
|\mu_{_H}|=\Lambda_{_{\rm NP}}$, respectively. In Fig.\ref{fig8}(a), the resonance around
$\Lambda_{_{\rm NP}}=200\;{\rm GeV}$ is arisen by the intervention
between the standard and supersymmetric fields.
As $\Lambda_{_{\rm NP}}\le350\;{\rm GeV}$, the difference between the
theoretical predictions on $d_e$ with and without $|\mu_{_H}|=|m_1|=|m_2|$
is not very obvious. With the increasing of $\Lambda_{_{\rm NP}}$,
the suppression of supersymmetric correction to $d_e$ without the assumption
is more stronger than that with the assumption.
A similar case exists in the 2-loop electroweak correction to $a_\mu$,
the suppression of supersymmetric correction to $a_\mu$ without the assumption
is stronger than that with the assumption when the energy scale $\Lambda_{_{\rm NP}}$ increases.

\section{Conclusions\label{sec4}}
\indent\indent
In this work, we analyzed the two-loop
supersymmetric corrections to the muon MDM and electron EDM
by the effective Lagrangian method in split scenarios.
In the concrete calculation, we keep
all dimension 6 operators. The ultraviolet divergence
caused by the divergent sub-diagrams is removed in the on-shell renormalization
schemes. After applying the equations of motion to the external leptons, we derive the
muon MDM and the electron EDM. Numerically, we analyze the dependence of the muon
MDM $a_\mu$ as well as the electron EDM $d_e$ on supersymmetric CP violating phases.
Adopting our assumptions on parameter space of the split supersymmetry,
we find that the correction from those two-loop diagrams to
$a_\mu$ is below $10^{-10}$ roughly for CP conservation, which is less than the
present experimental precision in magnitude. In other words,
the present experimental data do not put a very restrictive bound on
parameter space of split supersymmetry. Additional, the contribution to $d_\mu$
from this sector is sizable enough to be experimentally detected with
the experimental precision of near future.

\begin{acknowledgments}
\indent\indent
The work has been supported by the National Natural Science Foundation of China (NNSFC)
with No. 10675027. One of us (TFF) also acknowledges
the support from the ABRL Grant No. R14-2003-012-01001-0 of Korea at the early stage
of this work.
\end{acknowledgments}
\vspace{1.6cm}
\appendix

\section{The functions\label{ap1}}
\indent\indent
We list the tedious expressions of the functions adopted in the text
\begin{eqnarray}
&&\varrho_{_{i,j}}(x,y)={x^i\ln^jx-y^i\ln^jy\over x-y}\;,
\nonumber\\
&&\Omega_{_n}(x,y;u,v)={x^n\Phi(x,u,v)-y^n\Phi(y,u,v)\over x-y}\;,
\nonumber\\
&&T_1(x_1,x_2,x_3)={1\over x_1}\Bigg\{-4(2+\ln x_2)(\ln x_1-1)
-{\partial\over\partial x_3}\Big[\Big(1+2{x_2-x_3\over x_1}\Big)\Phi\Big]
(x_1,x_2,x_3)
\nonumber\\
&&\hspace{2.8cm}
+{\partial\over\partial x_3}\Big[\Big(1+2{x_2-x_3\over x_1}\Big)
\varphi_0+2(x_2-x_3)\varphi_1\Big](x_2,x_3)\Bigg\}\;,
\nonumber\\
&&T_2(x_1,x_2,x_3)
={1\over x_1}\Bigg[{\partial \Phi\over\partial x_3}
(x_1,x_2,x_3)-{\partial\varphi_0\over\partial x_3}(x_2,x_3)\Bigg]\;,
\nonumber\\
&&T_3(x_1,x_2,x_3)
=-{2\over x_1}(2+\ln x_3)+{2\over x_1}{\partial^2\over\partial x_3^2}
\Big(x_3\Phi\Big)(x_1,x_2,x_3)
\nonumber\\
&&\hspace{2.8cm}
-{2\over x_1}{\partial^2\over\partial x_3^2}
\Big(x_3\varphi_0\Big)(x_2,x_3)
-{4\over x_1}{\partial\Phi\over\partial x_3}
(x_1,x_2,x_3)
\nonumber\\
&&\hspace{2.8cm}
+{4\over x_1}{\partial\varphi_0\over\partial x_3}
(x_2,x_3)+{\partial^2\over\partial x_1\partial x_3}
\Big({x_2-x_3\over x_1}\varphi_0\Big)(x_2,x_3)
\nonumber\\
&&\hspace{2.8cm}
+{\partial^2\over\partial x_1\partial x_3}
\Big[\Big(1-{x_2-x_3\over x_1}\Big)\Phi\Big](x_1,x_2,x_3)\;,
\nonumber\\
&&T_4(x_1,x_2,x_3)={2\over x_1}\ln x_3-{2\over x_1^2}\Big(x_2-x_2\ln x_2
-x_3+x_3\ln x_3\Big)
\nonumber\\
&&\hspace{2.8cm}
-{\partial^3\over\partial x_1\partial x_3^2}\Big[{x_2x_3-x_3^2\over x_1}
\Big(\Phi(x_1,x_2,x_3)-\varphi_0(x_2,x_3)\Big)\Big]
\nonumber\\
&&\hspace{2.8cm}
+{1\over2}{\partial^3\over\partial x_1^2\partial x_3}
\Big[(x_2-3x_3-x_1)\Phi(x_1,x_2,x_3)\Big]
\nonumber\\
&&\hspace{2.8cm}
-{1\over2}{\partial^2\over\partial x_1\partial x_3}\Big[\Phi(x_1,x_2,x_3)
-{5\over x_1}(x_2-x_3)\Big(\Phi(x_1,x_2,x_3)
\nonumber\\
&&\hspace{2.8cm}
-\varphi_0(x_2,x_3)\Big)\Big]-{\partial^2\over\partial x_1^2}
\Big[{x_2-x_3\over x_1}\Big(\Phi(x_1,x_2,x_3)-\varphi_0(x_2,x_3)\Big)
\nonumber\\
&&\hspace{2.8cm}
+2\Phi(x_1,x_2,x_3)\Big]\;,
\nonumber\\
&&T_5(x_1,x_2,x_3)
={5\over12x_1}+\Big({5\over12x_1^2}+{\ln x_1\over3x_1^2}\Big)(x_2+x_3)
\nonumber\\
&&\hspace{2.8cm}
+\Big({7\over6x_1^2}+{2\over3x_1^2}\ln x_1\Big)(x_2\ln x_2+x_3\ln x_3)
\nonumber\\
&&\hspace{2.8cm}
+\Big({2\over3x_1^3}-{4\over3x_1^3}\ln x_1\Big)(x_2-x_3)^2
(1+\varrho_{_{1,1}}(x_2,x_3))
\nonumber\\
&&\hspace{2.8cm}
+{23\over6x_1^2}(x_2+x_3)\Big(1+\varrho_{_{1,1}}(x_2,x_3)\Big)
-{5\varrho_{_{2,1}}(x_2,x_3)\over x_1^2}
\nonumber\\
&&\hspace{2.8cm}
-{1\over3x_1^2}\Big(1-{2(x_2+x_3)\over x_1}\Big)\Big(\Phi(x_1,x_2,x_3)
-\varphi_0(x_2,x_3)\Big)
\nonumber\\
&&\hspace{2.8cm}
+{1\over3x_1}\Big({x_2+x_3\over x_1}-{2(x_2-x_3)^2\over x_1^2}\Big)\varphi_1(x_2,x_3)
\nonumber\\
&&\hspace{2.8cm}
+{1\over3x_1}\Big(1-{3(x_2+x_3)\over x_1}+{2(x_2-x_3)^2\over x_1^2}\Big)
{\partial\Phi\over\partial x_1}(x_1,x_2,x_3)
\nonumber\\
&&\hspace{2.8cm}
-{1\over3}\Big(1-{2(x_2+x_3)\over x_1}+{(x_2-x_3)^2\over x_1^2}\Big)
{\partial^2\Phi\over\partial x_1^2}(x_1,x_2,x_3)
\nonumber\\
&&\hspace{2.8cm}
-{(x_2-x_3)^2\over3x_1^2}\varphi_2(x_2,x_3)\;,
\nonumber\\
&&T_{6}(x_1,x_2,x_3)
=-{1\over x_1^2}\Big(\varphi_0-(x_2-x_3)
{\partial\varphi_0\over\partial x_3}\Big)(x_2,x_3)
+\Big[2x_3{\partial^3\Phi\over\partial x_1\partial x_3^2}
+{\partial^2\Phi\over\partial x_1^2}
\nonumber\\
&&\hspace{3.0cm}
+(x_1-x_2+x_3){\partial^3\Phi\over\partial x_1^2\partial x_3}
+{\Phi\over x_1^2}-{x_2-x_3\over x_1^2}{\partial\Phi\over\partial x_3}
-{1\over x_1}{\partial\Phi\over\partial x_1}
\nonumber\\
&&\hspace{3.0cm}
+(1+{x_2-x_3\over x_1}){\partial^2\Phi\over\partial x_1\partial x_3}\Big]
(x_1,x_2,x_3)\;,
\nonumber\\
&&T_{7}(x_1,x_2,x_3)
=-2{\partial^3\Phi\over\partial x_1^2\partial x_3}(x_1,x_2,x_3)
+{2\over x_1x_3}-{2\over x_1^2}\Big(\ln x_2-\ln x_3\Big)
\nonumber\\
&&\hspace{3.0cm}
+\Big({\partial^3\over\partial x_1^2\partial x_3}
-{\partial^3\over\partial x_1\partial x_3^2}
+{\partial^3\over\partial x_1^2\partial x_2}
+{\partial^3\over\partial x_1\partial x_2\partial x_3}\Big)\Big[\Phi(x_1,x_2,x_3)
\nonumber\\
&&\hspace{3.0cm}
-{x_2-x_3\over x_1}\Big(\Phi(x_1,x_2,x_3)-\varphi_0(x_2,x_3)\Big)\Big]\;,
\nonumber\\
&&T_{8}(x_1,x_2,x_3)
=-4\Big({\partial^3\Phi\over\partial x_1^2\partial x_3}
+{\partial^3\Phi\over\partial x_1^2\partial x_2}\Big)(x_1,x_2,x_3)
+{4\over x_1x_3}+{2\over x_1^2}(2+\ln x_2)
\nonumber\\
&&\hspace{3.0cm}
+\Big(2{\partial^3\over\partial x_1\partial x_3^2}
+{\partial^3\over\partial x_1^2\partial x_2}\Big)
\Big[{x_2-x_3\over x_1}\Big(\Phi(x_1,x_2,x_3)-\varphi_0(x_2,x_3)\Big)
\nonumber\\
&&\hspace{3.0cm}
-\Phi(x_1,x_2,x_3)\Big]\;,
\nonumber\\
&&T_{9}(x_1,x_2,x_3)={2\over x_1}\ln x_3
-{4x_3\over x_1^2}\Big({\partial\Phi\over\partial x_3}(x_1,x_2,x_3)
-{\partial\varphi_0\over\partial x_3}(x_2,x_3)\Big)
\nonumber\\
&&\hspace{3.0cm}
+{\partial^2\over\partial x_1\partial x_3}\Big((x_2-x_3)
{\Phi(x_1,x_2,x_3)-\varphi_0(x_2,x_3)\over x_1}-\Phi(x_1,x_2,x_3)\Big)
\nonumber\\
&&\hspace{3.0cm}
+{4\over x_1}\Big({\partial\Phi\over\partial x_3}
-{\partial\Phi\over\partial x_1}\Big)(x_1,x_2,x_3)
+{4x_3\over x_1}{\partial^2\Phi\over\partial x_1\partial x_3}(x_1,x_2,x_3)\;,
\nonumber\\
&&T_{10}(x_1,x_2,x_3)
={26\over x_1}+{17x_2\over x_1^2}+{29x_3\over x_1^2}
+{10\over x_1^2}\varrho_{_{2,1}}(x_2,x_3)
-{16(x_2-x_3)^2\over x_1^3}
\nonumber\\
&&\hspace{3.0cm}
-{10(x_2+x_3)\over x_1^2}\ln x_1-{6\ln x_3\over x_1}
+\Big[14-{16(x_2-x_3)\over x_1}
\Big]{x_2\ln x_2\over x_1^2}
\nonumber\\
&&\hspace{3.0cm}
+\Big[-4+{16(x_2-x_3)\over x_1}\Big]{x_3\ln x_3\over x_1^2}
+\Big[(x_2-x_3)^2-x_1^2\Big]{\partial^4\Phi\over\partial x_1^4}(x_1,x_2,x_3)
\nonumber\\
&&\hspace{3.0cm}
+\Big[-5x_1+6x_2+{3(x_2-x_3)^2\over x_1}\Big]
{\partial^3\Phi\over\partial x_1^3}(x_1,x_2,x_3)
\nonumber\\
&&\hspace{3.0cm}
+\Big[-{9(x_2-x_3)^2\over x_1^2}+{6x_2\over x_1}
+{3x_3\over x_1}\Big]{\partial^2\Phi\over\partial x_1^2}(x_1,x_2,x_3)
\nonumber\\
&&\hspace{3.0cm}
+\Big[-{12x_2\over x_1^2}-{6x_3\over x_1^2}+{18(x_2-x_3)^2\over x_1^3}\Big]
{\partial\Phi\over\partial x_1}(x_1,x_2,x_3)
\nonumber\\
&&\hspace{3.0cm}
+\Big[{12x_2\over x_1^3}+{6x_3\over x_1^3}-{18(x_2-x_3)^2\over x_1^4}\Big]
\Big(\Phi(x_1,x_2,x_3)-\varphi_0(x_2,x_3)\Big)
\nonumber\\
&&\hspace{3.0cm}
+{2x_3^2(x_2-x_3)\over x_1^2}\Big[{\partial^3\Phi\over\partial x_3^3}
(x_1,x_2,x_3)-{\partial^3\varphi_0\over\partial x_3^3}(x_2,x_3)\Big]
\nonumber\\
&&\hspace{3.0cm}
+\Big[{3x_\alpha x_\beta\over x_1^2}-{9x_\beta^2\over x_1^2}\Big]
\Big[{\partial^2\Phi\over\partial x_3^2}(x_1,x_2,x_3)
-{\partial^2\varphi_0\over\partial x_3^2}(x_2,x_3)\Big]
\nonumber\\
&&\hspace{3.0cm}
-\Big[{3x_\alpha\over x_1^2}+{9x_\beta\over x_1^2}+{18x_3(x_2-x_3)
\over x_1^3}\Big]\Big[{\partial\Phi\over\partial x_3}(x_1,x_2,x_3)
\nonumber\\
&&\hspace{3.0cm}
-{\partial\varphi_0\over\partial x_3}(x_2,x_3)\Big]-6x_3(x_2-x_3+x_1)
{\partial^4\Phi\over\partial x_1^3\partial x_3}(x_1,x_2,x_3)
\nonumber\\
&&\hspace{3.0cm}
+6x_3(x_2+x_3-x_1){\partial^4\Phi\over\partial x_1^2\partial x_3^2}
(x_1,x_2,x_3)
\nonumber\\
&&\hspace{3.0cm}
-2x_3^2\Big(1+{x_2-x_3\over x_1}\Big)
{\partial^4\Phi\over\partial x_1\partial x_3^3}(x_1,x_2,x_3)
\nonumber\\
&&\hspace{3.0cm}
+\Big[3x_1-3x_2-18x_3-{9x_3(x_2-x_3)\over x_1}\Big]
{\partial^3\Phi\over\partial x_1^2\partial x_3}(x_1,x_2,x_3)
\nonumber\\
&&\hspace{3.0cm}
+\Big[-21x_3-{3x_2x_3\over x_1}+{9x_3^2\over x_1}\Big]
{\partial^3\Phi\over\partial x_1\partial x_3^2}(x_1,x_2,x_3)
\nonumber\\
&&\hspace{3.0cm}
-\Big[6-{12x_2\over x_1}+{6x_3\over x_1}-{18x_3(x_2-x_3)\over x_1^2}\Big]
{\partial^2\Phi\over\partial x_1\partial x_3}(x_1,x_2,x_3)\;,
\nonumber\\
&&T_{11}(x_1,x_2,x_3)={2\ln x_3\over x_1}-{4(x_2-x_3)\over x_1^2}
-{4(x_2\ln x_2-x_3\ln x_3)\over x_1^2}
\nonumber\\
&&\hspace{3.0cm}
-{4(x_2-x_3)\over x_1^3}\Big(\Phi(x_1,x_2,x_3)-\varphi_0(x_2,x_3)\Big)
+{4(x_2-x_3)\over x_1^2}{\partial\Phi\over\partial x_1}(x_1,x_2,x_3)
\nonumber\\
&&\hspace{3.0cm}
-\Big(1+{2(x_2-x_3)\over x_1}\Big){\partial^2\Phi\over\partial x_1^2}
(x_1,x_2,x_3)
-{2x_3\over x_1^2}\Big({\partial\Phi\over\partial x_3}(x_1,x_2,x_3)
\nonumber\\
&&\hspace{3.0cm}
-{\partial\varphi_0\over\partial x_3}(x_2,x_3)\Big)+{x_3(x_2-x_3)\over x_1^2}
\Big({\partial^2\Phi\over\partial x_3^2}(x_1,x_2,x_3)
-{\partial^2\varphi_0\over\partial x_3^2}(x_2,x_3)\Big)
\nonumber\\
&&\hspace{3.0cm}
-2{\partial^2\Phi\over\partial x_1\partial x_3}(x_1,x_2,x_3)
-x_3\Big(1+{x_2-x_3\over x_1}\Big)
{\partial^3\Phi\over\partial x_1\partial x_3^2}(x_1,x_2,x_3)
\nonumber\\
&&\hspace{3.0cm}
+\Big(x_2+x_3-x_1\Big){\partial^3\Phi\over\partial x_1^2\partial x_3}
(x_1,x_2,x_3)\;,
\nonumber\\
&&T_{12}(x_1,x_2,x_3)
=-{52\over x_1^2}+{4\over x_1x_3}+{20\over x_1^2}\ln x_1-{18\ln x_3\over x_1^2}
-{20\over x_1^2}\varrho_{_{1,1}}(x_2,x_3)
\nonumber\\
&&\hspace{3.0cm}
-{12\over x_1^3}\Big(\Phi(x_1,x_2,x_3)-\varphi_0(x_2,x_3)\Big)
+{12\over x_1^2}{\partial\Phi\over\partial x_1}(x_1,x_2,x_3)
\nonumber\\
&&\hspace{3.0cm}
-{6\over x_1}{\partial^2\Phi\over\partial x_1^2}(x_1,x_2,x_3)
-\Big(17{\partial^3\Phi\over\partial x_1^3}
+2x_1{\partial^4\Phi\over\partial x_1^4}\Big)(x_1,x_2,x_3)
\nonumber\\
&&\hspace{3.0cm}
+{6\over x_1^2}\Big(1+{2(x_2-x_3)\over x_1}\Big)
\Big({\partial\Phi\over\partial x_3}(x_1,x_2,x_3)
-{\partial\varphi_0\over\partial x_3}(x_2,x_3)\Big)
\nonumber\\
&&\hspace{3.0cm}
-{3(x_2-2x_3)\over x_1^2}\Big({\partial^2\Phi\over\partial x_3^2}
(x_1,x_2,x_3)-{\partial^2\varphi_0\over\partial x_3^2}(x_2,x_3)\Big)
\nonumber\\
&&\hspace{3.0cm}
-{x_3(x_2-x_3)\over x_1^2}\Big({\partial^3\Phi\over\partial x_3^3}
(x_1,x_2,x_3)-{\partial^3\varphi_0\over\partial x_3^3}(x_2,x_3)\Big)
\nonumber\\
&&\hspace{3.0cm}
-x_3\Big(1-{x_2-x_3\over x_1}\Big){\partial^4\Phi\over\partial x_1\partial x_3^3}
(x_1,x_2,x_3)
\nonumber\\
&&\hspace{3.0cm}
-{6\over x_1}\Big(1+{2(x_2-x_3)\over x_1}\Big)
{\partial^2\Phi\over\partial x_1\partial x_3}(x_1,x_2,x_3)
\nonumber\\
&&\hspace{3.0cm}
-\Big[3\Big(1-{x_2-2x_3\over x_1}\Big){\partial^3\Phi\over\partial x_1\partial x_3^2}
+6\Big(2-{x_2-x_3\over x_1}\Big){\partial^3\Phi\over\partial x_1^2\partial x_3}\Big]
(x_1,x_2,x_3)
\nonumber\\
&&\hspace{3.0cm}
+3(x_2-x_3-x_1){\partial^4\Phi\over\partial x_1^3\partial x_3}
(x_1,x_2,x_3)-6{\partial^4\Phi\over\partial x_1^2\partial x_3^2}(x_1,x_2,x_3)\;,
\nonumber\\
&&T_{13}(x_1,x_2,x_3)
={1\over x_1x_3}+{2\over x_1^2}\Big({\partial\Phi\over\partial x_3}
(x_1,x_2,x_3)-{\partial\varphi_0\over\partial x_3}(x_2,x_3)\Big)
-{2\over x_1}{\partial^2\Phi\over\partial x_1\partial x_3}(x_1,x_2,x_3)
\nonumber\\
&&\hspace{3.0cm}
-{x_2-x_3\over x_1^2}
\Big({\partial^2\Phi\over\partial x_3^2}
(x_1,x_2,x_3)
-{\partial^2\varphi_0\over\partial x_3^2}(x_2,x_3)\Big)
\nonumber\\
&&\hspace{3.0cm}
-\Big(1-{x_2-x_3\over x_1}\Big)
{\partial^3\Phi\over\partial x_1\partial x_3^2}(x_1,x_2,x_3)
-2{\partial^3\Phi\over\partial x_1^2\partial x_3}(x_1,x_2,x_3)\;,
\nonumber\\
&&F_1(x_1,x_2,x_3,x_4)={1\over x_1x_2}{\partial\over\partial x_4}\Big((x_3-x_4)\varphi_0\Big)(x_3,x_4)
\nonumber\\
&&\hspace{3.6cm}
+{1\over x_1-x_2}\Big\{{\partial\over\partial
x_4}\Big[\Big(1+{x_3-x_4\over x_1}\Big)\Phi\Big](x_1,x_3,x_4)
\nonumber\\
&&\hspace{3.6cm}
-{\partial\over\partial x_4}\Big[\Big(1+{x_3-x_4\over x_2}\Big)\Phi\Big](x_2,x_3,x_4)\Big\}\;,
\nonumber\\
&&F_2(x_1,x_2,x_3,x_4)=-{1\over x_1x_2}{\partial\over\partial x_4}\Big((x_3-x_4)\varphi_0\Big)(x_3,x_4)
\nonumber\\
&&\hspace{3.6cm}
+{1\over x_1-x_2}\Big\{{\partial\over\partial
x_4}\Big[\Big(1-{x_3-x_4\over x_1}\Big)\Phi\Big](x_1,x_3,x_4)
\nonumber\\
&&\hspace{3.6cm}
-{\partial\over\partial x_4}\Big[\Big(1-{x_3-x_4\over x_2}\Big)\Phi\Big](x_2,x_3,x_4)\Big\}\;,
\nonumber\\
&&F_3(x_1,x_2,x_3,x_4)=
2(\ln x_4-1)\varrho_{_{0,1}}(x_1,x_2)-{6(x_3-x_4)\over x_1x_2}
-{6(x_3\ln x_3-x_4\ln x_4)\over x_1x_2}
\nonumber\\
&&\hspace{3.6cm}
+{x_1x_2+2(x_1+x_2)(x_3-x_4)\over x_1^2x_2^2}\varphi_0(x_3,x_4)
-{x_3-3x_4\over x_1x_2}{\partial\varphi_0\over\partial x_4}(x_3,x_4)
\nonumber\\
&&\hspace{3.6cm}
-{x_4(x_3-x_4)\over x_1x_2}
{\partial^2\varphi_0\over\partial x_4^2}(x_3,x_4)
-\Big({\partial\over\partial x_4}+x_4{\partial^2\over\partial x_4^2}\Big)
\Omega_{_0}(x_1,x_2;x_3,x_4)
\nonumber\\
&&\hspace{3.6cm}
+\Big(1-(x_3-3x_4){\partial\over\partial x_4}-x_4
(x_3-x_4){\partial^2\over\partial x_4^2}\Big)\Omega_{_{-1}}(x_1,x_2;x_3,x_4)
\nonumber\\
&&\hspace{3.6cm}
-\Big({\partial\over\partial x_1}+{\partial\over\partial x_2}\Big)^2
\Big[\Omega_{_1}(x_1,x_2;x_3,x_4)
+(x_3-x_4)\Omega_{_0}(x_1,x_2;x_3,x_4)\Big]
\nonumber\\
&&\hspace{3.6cm}
-2\Big({\partial\over\partial x_1}+{\partial\over\partial x_2}\Big)
\Big[{\partial\Omega_{_1}\over\partial x_4}(x_1,x_2;x_3,x_4)
-(x_3+x_4){\partial\Omega_{_0}\over\partial x_4}(x_1,x_2;x_3,x_4)\Big]
\nonumber\\
&&\hspace{3.6cm}
-2(x_3-x_4)\Big({\partial\over\partial x_1}+{\partial\over\partial x_2}\Big)
\Omega_{_{-1}}(x_1,x_2;x_3,x_4)\;,
\nonumber\\
&&F_4(x_1,x_2,x_3,x_4)=
2(\ln x_4-1)\varrho_{_{0,1}}(x_1,x_2)-{6(x_3-x_4)\over x_1x_2}
-{6(x_3\ln x_3-x_4\ln x_4)\over x_1x_2}
\nonumber\\
&&\hspace{3.6cm}
-{x_1x_2-2(x_1+x_2)(x_3-x_4)\over x_1^2x_2^2}\varphi_0(x_3,x_4)
+{x_3+x_4\over x_1x_2}{\partial\varphi_0\over\partial x_4}(x_3,x_4)
\nonumber\\
&&\hspace{3.6cm}
-{x_4(x_3-x_4)\over x_1x_2}
{\partial^2\varphi_0\over\partial x_4^2}(x_3,x_4)
+\Big(-{\partial\over\partial x_4}+x_4{\partial^2\over\partial x_4^2}\Big)
\Omega_{_0}(x_1,x_2;x_3,x_4)
\nonumber\\
&&\hspace{3.6cm}
+\Big(-1+(x_3+x_4){\partial\over\partial x_4}-x_4
(x_3-x_4){\partial^2\over\partial x_4^2}\Big)\Omega_{_{-1}}(x_1,x_2;x_3,x_4)
\nonumber\\
&&\hspace{3.6cm}
+\Big({\partial\over\partial x_1}+{\partial\over\partial x_2}\Big)^2
\Big[\Omega_{_1}(x_1,x_2;x_3,x_4)
-(x_3-x_4)\Omega_{_0}(x_1,x_2;x_3,x_4)\Big]
\nonumber\\
&&\hspace{3.6cm}
-2\Big({\partial\over\partial x_1}+{\partial\over\partial x_2}\Big)
\Big[\Omega_{_0}(x_1,x_2;x_3,x_4)
-2x_4{\partial\Omega_{_0}\over\partial x_4}(x_1,x_2;x_3,x_4)\Big]
\nonumber\\
&&\hspace{3.6cm}
-2(x_3-x_4)\Big({\partial\over\partial x_1}+{\partial\over\partial x_2}\Big)
\Omega_{_{-1}}(x_1,x_2;x_3,x_4)\;,
\nonumber\\
&&F_5(x_1,x_2,x_3,x_4)=-2(2+\ln x_4)\varrho_{_{0,1}}(x_1,x_2)
+{1\over x_1x_2}\varphi_0(x_3,x_4)
\nonumber\\
&&\hspace{3.6cm}
-{x_3-x_4\over x_1x_2}{\partial\varphi_0\over\partial x_4}(x_3,x_4)
-{\partial\Omega_{_0}\over\partial x_4}(x_1,x_2;x_3,x_4)
\nonumber\\
&&\hspace{3.6cm}
+\Big(1-(x_3-x_4){\partial\over\partial x_4}\Big)\Omega_{-1}
(x_1,x_2;x_3,x_4)\;,
\nonumber\\
&&F_6(x_1,x_2,x_3,x_4)=2(2+\ln x_4)\varrho_{_{0,1}}(x_1,x_2)
-{1\over x_1x_2}\varphi_0(x_3,x_4)
\nonumber\\
&&\hspace{3.6cm}
+{x_3-x_4\over x_1x_2}{\partial\varphi_0\over\partial x_4}(x_3,x_4)
-{\partial\Omega_{_0}\over\partial x_4}(x_1,x_2;x_3,x_4)
\nonumber\\
&&\hspace{3.6cm}
-\Big(1-(x_3-x_4){\partial\over\partial x_4}\Big)\Omega_{-1}
(x_1,x_2;x_3,x_4)\;.
\label{funs}
\end{eqnarray}

The concrete expression of $\Phi(x,y,z)$ can be found in \cite{Feng2,2vac}. In the limit
$z\ll x,y$, we can expand $\Phi(x,y,z)$ according $z$ as
\begin{eqnarray}
&&\Phi(x,y,z)=\varphi_0(x,y)+z\varphi_1(x,y)+{z^2\over2!}\varphi_2(x,y)
+{z^3\over3!}\varphi_3(x,y)+{z^4\over4!}\varphi_4(x,y)
\nonumber\\
&&\hspace{2.2cm}
+2z\Big(\ln z-1\Big)\Big(1+\varrho_{_{1,1}}(x,y)\Big)
\nonumber\\
&&\hspace{2.2cm}
-2z^2\Big({\ln z\over2!}-{3\over4}\Big)
\Big({x+y\over(x-y)^2}+{2xy\over(x-y)^3}\ln{y\over x}\Big)
\nonumber\\
&&\hspace{2.2cm}
-{2z^3\over(x-y)^2}\Big({\ln z\over3!}-{11\over36}\Big)\Big(1+{12xy\over(x-y)^2}
+{6xy(x+y)\over(x-y)^3}\ln{y\over x}\Big)
\nonumber\\
&&\hspace{2.2cm}
-2z^4\Big({\ln z\over4!}-{25\over288}\Big)\Big(
{2x^3+58x^2y+58xy^2+2y^3\over(x-y)^6}
\nonumber\\
&&\hspace{2.2cm}
+{24xy(x^2+3xy+y^2)\over(x-y)^7}\ln{y\over x}\Big)+\cdots
\label{phi-expand}
\end{eqnarray}
with
\begin{eqnarray}
&&\varphi_0(x,y)=\left\{\begin{array}{ll}(x+y)\ln x\ln y+(x-y)\Theta(x,y)
\;,&x>y\;;\\
2x\ln^2x\;,&x=y\;;\\
(x+y)\ln x\ln y+(y-x)\Theta(y,x)
\;,&x<y\;.\end{array}\right.
\label{varphi0}
\end{eqnarray}

\begin{eqnarray}
&&\varphi_1(x,y)=\left\{\begin{array}{ll}-\ln x\ln y-{x+y\over x-y}\Theta(x,y)
\;,&x>y\;;\\
4-2\ln x-\ln^2x\;,&x=y\;;\\
-\ln x\ln y-{x+y\over y-x}\Theta(y,x)
\;,&x<y\;.\end{array}\right.
\label{varphi1}
\end{eqnarray}

\begin{eqnarray}
&&\varphi_2(x,y)=\left\{\begin{array}{ll}
{(2x^2+6xy)\ln x-(6xy+2y^2)\ln y\over(x-y)^3}-{4xy\over(x-y)^3}\Theta(x,y)
\;,&x>y\;;\\
-{5\over9x}+{2\over3x}\ln x\;,&x=y\;;\\
{(2x^2+6xy)\ln x-(6xy+2y^2)\ln y\over(x-y)^3}-{4xy\over(y-x)^3}\Theta(y,x)
\;,&x<y\;.\end{array}\right.
\label{varphi2}
\end{eqnarray}

\begin{eqnarray}
&&\varphi_3(x,y)=\left\{\begin{array}{ll}-{12xy(x+y)\over(x-y)^5}\Theta(x,y)
-{2(x^2+xy+y^2)\over(x-y)^4} & \\
+{2(x^3+14x^2y+11xy^2)\ln x-2(y^3+14xy^2+11x^2y)\ln y\over(x-y)^5}
\;,&x>y\;;\\
-{53\over150x^2}+{1\over5x^2}\ln x\;,&x=y\;;\\
-{12xy(x+y)\over(y-x)^5}\Theta(y,x)-{2(x^2+xy+y^2)\over(x-y)^4} & \\
+{2(x^3+14x^2y+11xy^2)\ln x-2(y^3+14xy^2+11x^2y)\ln y\over(x-y)^5}
\;,&x<y\;.
\end{array}\right.
\label{varphi3}
\end{eqnarray}

\begin{eqnarray}
&&\varphi_4(x,y)=\left\{\begin{array}{ll}
-{48xy(x^2+3xy+y^2)\over(x-y)^7}\Theta(x,y)-{2(3x^3+61x^2y+61xy^2+3y^3)\over(x-y)^6} & \\
+{4(x^4+3x^3y-45x^2y^2-25xy^3)\ln x-4(y^4+3y^3x-45x^2y^2-25yx^3)\ln y\over(x-y)^7}
\;,&x>y\;;\\
-{598\over2205x^3}+{1\over210x^3}\ln x\;,&x=y\;;\\
-{48xy(x^2+3xy+y^2)\over(x-y)^7}\Theta(y,x)
-{2(3x^3+61x^2y+61xy^2+3y^3)\over(x-y)^6}& \\
+{4(x^4+3x^3y-45x^2y^2-25xy^3)\ln x-4(y^4+3y^3x-45x^2y^2-25yx^3)\ln y\over(x-y)^7}
\;,&x<y\;.\end{array}\right.
\label{varphi4}
\end{eqnarray}

Here, the function $\Theta(x,y)$ is defined as
\begin{eqnarray}
&&\Theta(x,y)=\ln x\ln{y\over x}-2\ln(x-y)\ln{y\over x}-2Li_2({y\over x})+{\pi^2\over3}\;.
\label{theta}
\end{eqnarray}


\begin{thebibliography}{99}
\bibitem{exp}[The Muon $g-2$ Collaboration], Phys.~Rev.~Lett.~{\bf 92}(2004)161802.
\bibitem{Jegerlehner}J.~P.~Miller, E.~de~Rafael and B.~L.~Roberts,
{\it Muon (g-2): experiment and theory}, Rep.~Prog.~Phys.~{\bf 70}(2007)795;
F.~Jegerlehner, Acta~Phys.~Polon.~B~{\bf 38}(2007)3021.
\bibitem{sm1}M.~Davier, S.~Eidelman, A.~Hocker and Z.~Zhang, Eur.~Phys.
~J.~C {\bf 31}(2003)503.
\bibitem{sm2}K.~Hagiwara, A.~Martin, D.~Normura and T.~Teubner, Phys.
~Lett.~B~{\bf 557}(2003)69.
\bibitem{sm3}S.~Ghozzi, F.~Jegerlehner, Phys.~Lett.~B~{\bf 583}(2004)222;
M.~Passera, J.~Phys.~G~{\bf 31}(2005)R75;
Nucl.~Phys.~Proc.~Suppl.~{\bf 155}(2006)365;
M. Passera, {\it The calculation of the muon g-2 and
$\Delta(\alpha(M(Z)^2)$}, PoS HEP2005,~(2006)305.
\bibitem{sm-2l}A.~Czarnecki and W.~Marciano, Phys.~Rev.~D {\bf 64}(2001)013014;
M.~Knecht, Lect. Notes. Phys. {\bf 629}(2004)37.
\bibitem{Arkani1}N.~Arkani-Hamed and S.~Dimopoulos, JHEP~{\bf 073}(2005)0506.
\bibitem{Arkani2}N.~Arkani-Hamed et~al., Nucl.~Phys.~B {\bf 709}(2005)3.
\bibitem{czarnecki}A.~Czarnecki, B.~Krause and W.~J.~Marciano,
Phys.~Rev.~D {\bf 52}(1995)2619; Phys.~Rev.~Lett. {\bf 76},(1996)3267;
T.~Kukhto, E.~Kuraev, A.~Schiller and Z.~Silagadze, Nucl.~Phys.~B {\bf 371}(1992)567.
\bibitem{heinemeyer1}S.~Heinemeyer, D.~St\"ockinger and G.~Weiglein,
Nucl. Phys. B {\bf 690}(2004)62.
\bibitem{heinemeyer2}S.~Heinemeyer, D.~St\"ockinger and G.~Weiglein,
Nucl. Phys. B {\bf 699}(2004)103.
\bibitem{geng}C.~Chen, C.~Geng, Phys. Lett. B {\bf 511}(2001)77;
A.~Arhrib and S.~Baek, Phys.~Rev.~D {\bf 65}(2002)075002.
\bibitem{DChang}D.~Chang, W.~Chang, and W.~Keung, Phys.~Rev.~D {\bf 71}(2005)076006.
\bibitem{Giudice}G.~Giudice, A.~Romanino, Phys.~Lett.~B {\bf 634}(2005)69.
\bibitem{Feng1}T.-F. Feng, Phys.~Rev.~D~{\bf 70}(2004)096012.
\bibitem{Feng2}T.-F. Feng, X.-Q. Li, J. Maalampi, X.-M. Zhang, Phys.~Rev.~D {\bf 71}(2005)056005.
\bibitem{Feng3}T.-F. Feng, T. Huang, X.-Q. Li, X.-M. Zhang,
S.-M. Zhao, Phys.~Rev.~D~{\bf 68}(2003)016004; T.-F. Feng, X.-Q. Li,
L. Lin, J. Maalampi, and H.-S. Song, Phys.~Rev.~D {\bf 73}(2006)116001.
\bibitem{onshell}M.~Bohm, H.~Spiesberger, W.~Hollik,
Fortsch.~Phys. {\bf 34}(1986)687; A.~Denner, {\it ibid.}~{\bf 41}(1993)307.
\bibitem{nonlinear-R-xi}L.~F.~Abbott, Nucl.~Phys.~B~{\bf 185}(1981)189;
M.~B.~Gavela, G.~Girardi, C.~Malleville, and P.~Sorba,  Nucl.~Phys.~B~{\bf 193}(1981)257;
N.~G.~Deshpande, M.~Nazerimonfared, Nucl.~Phys.~B~{\bf 213}(1983)390.
\bibitem{nexp}Y.K. Semertzidis et al., {\it Sensitive search for a permanent muon electric dipole moment},
hep-ph/0012087.
\bibitem{Regan}B.~C.~Regan, E.~D.~Commins,~C.~J.~Schmidt and D.~DeMille,
Phys.~Rev.~Lett. {\bf 88}(2002)071805.
\bibitem{Kawell}D.~Kawell, F.~Bay, S.~Bickman, Y.~Jiang and D.~Demille,
AIP~Conf.~Proc. {\bf 698}(2004)192.
\bibitem{Barr-Zee}S.~M.~Barr and A.~Zee, Phys. Rev. Lett. {\bf 65}(1990)21.
\bibitem{BZ-mssm}D.~Chang, W.-Y.~Keung, and A.~Pilaftsis, Phys.~Rev.~Lett. {\bf 82}(1999)900;
{\bf 83}(1999)3972(E); A.~Pilaftsis, Phys.~Lett.~B {\bf 471}(1999)174;
D.~Chang, W.-F.~Chang, and W.-Y.~Keung, {\it ibid.}~{\bf 478}(2000)239;
A.~Pilaftsis, Nucl.~Phys.~B~{\bf 644}(2002)263.
\bibitem{Dress}M.~Dress, {\it Some comments on split supersymmetry}, hep-ph/0501106.
\bibitem{Pilaftsis}A.~Pilaftsis,~Phys.~Rev.~D~{\bf 58}(1998)096010;
Phys.~Lett.~B~{\bf 435}(1998)88;
A.~Pilaftsis, C.~E.~M.~Wagner, Nucl.~Phys.~B~{\bf 533}(1999)3;
M.~Carena, J.~Ellis, A.~Pilaftsis, C.~E.~M.~Wagner,~{\it ibid}.~{\bf 586}(2000)92;
{\it ibid}.~{\bf 625}(2002)345.
\bibitem{2vac}A. I. Davydychev and J. B. Tausk, Nucl. Phys. B.
{\bf 397}, 123(1993).
\end{thebibliography}
\end{document}